# Title Page

# SiCmiR Atlas: Single-Cell miRNA Landscapes Reveals Hub-miRNA and Network Signatures in Human Cancers

Xiao-Xuan Cai[1,2], Jing-Shan Liao[2], Jia-Jun Ma[2], Yu-Xuan Pang[1], Yi-Gang Chen[1,2], Yang-Chi-Dung Lin[1,2,3], Yi-Dan Chen[1,2], Xin Cao[2], Yi-Cheng Zhang[2], Tao-Sheng Xu[1], Tzong-Yi Lee[4], Hsi-Yuan Huang[1,2,3*], and Hsien-Da Huang[1,2,3,5*]

1 Warshel Institute for Computational Biology, School of Medicine, The Chinese University of Hong Kong, Shenzhen, Longgang District, Shenzhen, Guangdong, 518172, PR China

2 School of Medicine, The Chinese University of Hong Kong, Shenzhen, Longgang District, Shenzhen, Guangdong, 518172, PR China

3 Guangdong Provincial Key Laboratory of Digital Biology and Drug Development, The Chinese University of Hong Kong, Shenzhen, Guangdong, 518172, PR China

4 Institute of Bioinformatics and Systems Biology, National Yang Ming Chiao Tung University, 300093 Hsinchu, Taiwan

5 Department of Endocrinology, Key Laboratory of Endocrinology of National Ministry of Health, Peking Union Medical College Hospital, Chinese Academy of Medical Sciences & Peking Union Medical College, Beijing, 100730, PR China

*Correspondence authors: huanghsienda@cuhk.edu.cn (H.D.H), huanghsiyuan@cuhk.edu.cn (H.Y.H)
Contributing authors: xiaoxuancai@link.cuhk.edu.cn; jingshanliao@link.cuhk.edu.cn; jiajunma@link.cuhk.edu.cn; yuxuanpang@link.cuhk.edu.cn; yigangchen@link.cuhk.edu.cn; yangchidung@cuhk.edu.cn; yidanchen@link.cuhk.edu.cn; xincao@link.cuhk.edu.cn; yichengzhang@link.cuhk.edu.cn; taosheng.x@gmail.com; leetzongyi@nycu.edu.tw; huanghsiyuan@cuhk.edu.cn; huanghsienda@cuhk.edu.cn.

**Data availability statement**

The authors declare that data supporting the findings of this study are available within the article and supplementary information. Package for usage of SiCmiR can be retrieved at https://github.com/Cristinex/SiCmiR. Online application is deposited at https://awi.cuhk.edu.cn/~SiCmiR/. Data supporting this study are openly available in Zenodo repository at https://doi.org/10.5281/zenodo.16025109.

**Funding statement**

This research was funded by Shenzhen Science and Technology Innovation Program [JCYJ20220530143615035]; Guangdong S&T programme [2024A0505050001, 2024A0505050002]; Warshel Institute for Computational Biology funding from Shenzhen City and Longgang District [LGKCSDPT2024001]; Shenzhen-Hong Kong Cooperation Zone for Technology and Innovation [HZQB-KCZYB-2020056, P2-2022-HDH-001-A]; Guangdong Young Scholar Development Fund of Shenzhen Ganghong Group Co., Ltd. [2021E0005, 2022E0035, 2023E0012].

**Conflict of interest disclosure**

The authors have declared no competing interests.

**Ethics approval statement**

Not applicable.

**Patient consent statement**

Not applicable.

**Permission to reproduce material from other sources**

Not applicable.

**Clinical trial registration**

Not applicable.

# SiCmiR Atlas: Single-Cell miRNA Landscapes Reveals Hub-miRNA and Network Signatures in Human Cancers

Xiao-Xuan Cai[1,2], Jing-Shan Liao[2], Jia-Jun Ma[2], Yu-Xuan Pang[1], Yi-Gang Chen[1,2], Yang-Chi-Dung Lin[1,2,3], Yi-Dan Chen[1,2], Xin Cao[2], Yi-Cheng Zhang[2], Tao-Sheng Xu[1], Tzong-Yi Lee[4], Hsi-Yuan Huang[1,2,3*], and Hsien-Da Huang[1,2,3,5*]

1 Warshel Institute for Computational Biology, School of Medicine, The Chinese University of Hong Kong, Shenzhen, Longgang District, Shenzhen, Guangdong, 518172, PR China

2 School of Medicine, The Chinese University of Hong Kong, Shenzhen, Longgang District, Shenzhen, Guangdong, 518172, PR China

3 Guangdong Provincial Key Laboratory of Digital Biology and Drug Development, The Chinese University of Hong Kong, Shenzhen, Guangdong, 518172, PR China

4 Institute of Bioinformatics and Systems Biology, National Yang Ming Chiao Tung University, 300093 Hsinchu, Taiwan

5 Department of Endocrinology, Key Laboratory of Endocrinology of National Ministry of Health, Peking Union Medical College Hospital, Chinese Academy of Medical Sciences & Peking Union Medical College, Beijing, 100730, PR China

*Correspondence authors: huanghsienda@cuhk.edu.cn (H.D.H), huanghsiyuan@cuhk.edu.cn (H.Y.H)

Contributing authors: xiaoxuancai@link.cuhk.edu.cn; jingshanliao@link.cuhk.edu.cn; jiajunma@link.cuhk.edu.cn; yuxuanpang@link.cuhk.edu.cn; yigangchen@link.cuhk.edu.cn; yangchidung@cuhk.edu.cn; yidanchen@link.cuhk.edu.cn; xincao@link.cuhk.edu.cn; yichengzhang@link.cuhk.edu.cn; taosheng.x@gmail.com; leetzongyi@nycu.edu.tw; huanghsiyuan@cuhk.edu.cn; huanghsienda@cuhk.edu.cn.

**Abstract**

microRNAs (miRNAs) are pivotal post-transcriptional regulators whose single-cell behaviour has remained largely inaccessible owing to technical barriers in single-cell small-RNA profiling. We present SiCmiR, a two-layer neural network that predicts miRNA expression profile from only 977 LINCS L1000 landmark genes reducing sensitivity to dropout of single-cell RNA-seq (scRNA-seq) data. Proof-of-concept analyses illustrate how SiCmiR can uncover candidate hub-miRNAs in bulk-seq cell lines and hepatocellular carcinoma, scRNA-seq pancreatic ductal carcinoma and ACTH-secreting pituitary adenoma and extracellular-vesicle (EV)-mediated crosstalk in glioblastoma. Trained on 6,462 TCGA paired miRNA–mRNA samples, SiCmiR attains state-of-the-art accuracy on held-out cancers and generalises to unseen cancer types, drug perturbations and scRNA-seq. We next constructed SiCmiR-Atlas, containing 362 public datasets, 9.36 million cells, 726 cell types, which is the first dedicated database of single-cell mature miRNA expression—providing interactive visualisation, biomarker identification and cell-type-resolved miRNA–target networks. SiCmiR transforms bulk-derived statistical power into a single-cell view of miRNA biology and provides a community resource SiCmiR Atlas for biomarker discovery. SiCmiR Atlas is available at https://awi.cuhk.edu.cn/~SiCmiR/.

**Introduction**

microRNAs (miRNAs), small non-coding RNAs regulating gene expression, predominantly act as repressors by binding to the 3'-untranslated region (3'UTR) of mRNAs, initiating mRNA degradation and blocking translation, though certain miRNAs have been reported to stabilize mRNA and to enhance its activity.[1] Dysregulation of miRNAs and their protein translation in the regulatory network underpins virtually every cancer hallmark—proliferation, stemness, invasion and immune evasion—making miRNAs valuable biomarkers and therapeutic entry points.[2] miRNAs that exhibit substantial associations with mRNAs actively participate in the regulatory network, highlighting their potential as hub-miRNAs.[3]

According to the records of miRBase, there are 2,656 mature miRNAs have been identified in humans,[4] although not all have been assigned functional significance. This gap in knowledge may be attributed to factors such as inadequate sample sizes, sequencing batch effects, and limitations in capturing the true miRNA landscape.[5] Additionally, existing approaches for identifying disease-related miRNAs, such as weighted gene co-expression network analysis (WGCNA), which detects co-expression patterns and hub nodes through topological overlap,[6] or annotation-based methods relying on miRNA-target databases,[7-9] suffer from incomplete functional annotations and limited coverage.[10] Machine learning and deep learning models have been introduced to predict disease-related miRNAs based on known miRNA-disease associations,[11, 12] yet challenges persist due to tumor heterogeneity,[13] which complicates hub-miRNA discovery.

While single-cell RNA-sequencing (scRNA-seq) has significantly advanced mRNA-level investigations, single-cell miRNA sequencing remains in its early stages, with limited applications.[14-17] Current single-cell miRNA profiling techniques face several obstacles, including dependence on polyadenylation, adaptor dimer formation, high data sparsity, challenges in distinguishing miRNAs from other small non-coding RNAs, and inconsistencies in protocol reproducibility.[18] These technical limitations have hindered the identification of hub-miRNAs in cancers, emphasizing the need for improved methodologies. Recent efforts such as miRSCAPE[19] and miTEA-HiRes[20] have markedly improved the recovery of miRNA activity at single-cell resolution. miRSCAPE requires around 20,000 gene features and therefore suffers from zero inflation in scRNA-seq; miTEA-HiRes infers miRNA activity by testing target-gene enrichment within spatial transcriptomic spots, a procedure that depends on canonical target lists—thus failing to capture continuous miRNA expression profile.

SiCmiR addresses these limitations by (i) relying on just 977 landmark genes, (ii) supporting single-cell inputs without pre-clustering. SiCmiR shows its robustness in hepatocellular carcinoma, glioblastoma and ACTH-secreting pituitary adenoma and revealed 414 hub-miRNAs in cancer and also extracellular-vesicle (EV) mediated intercellular communication at single-cell resolution. To maximize the method's utility, we further compiled the predictions across multiple datasets as SiCmiR-Atlas, an openly searchable database, the first public database dedicated to single-cell mature miRNA expression, providing interactive visualization, biomarker mining and cell-type-resolved miRNA–target networks. By coupling bulk-derived statistical power with single-cell granularity, SiCmiR establishes a practical route to dissect miRNA regulation in heterogeneous tissues and accelerates biomarker discovery in oncology.

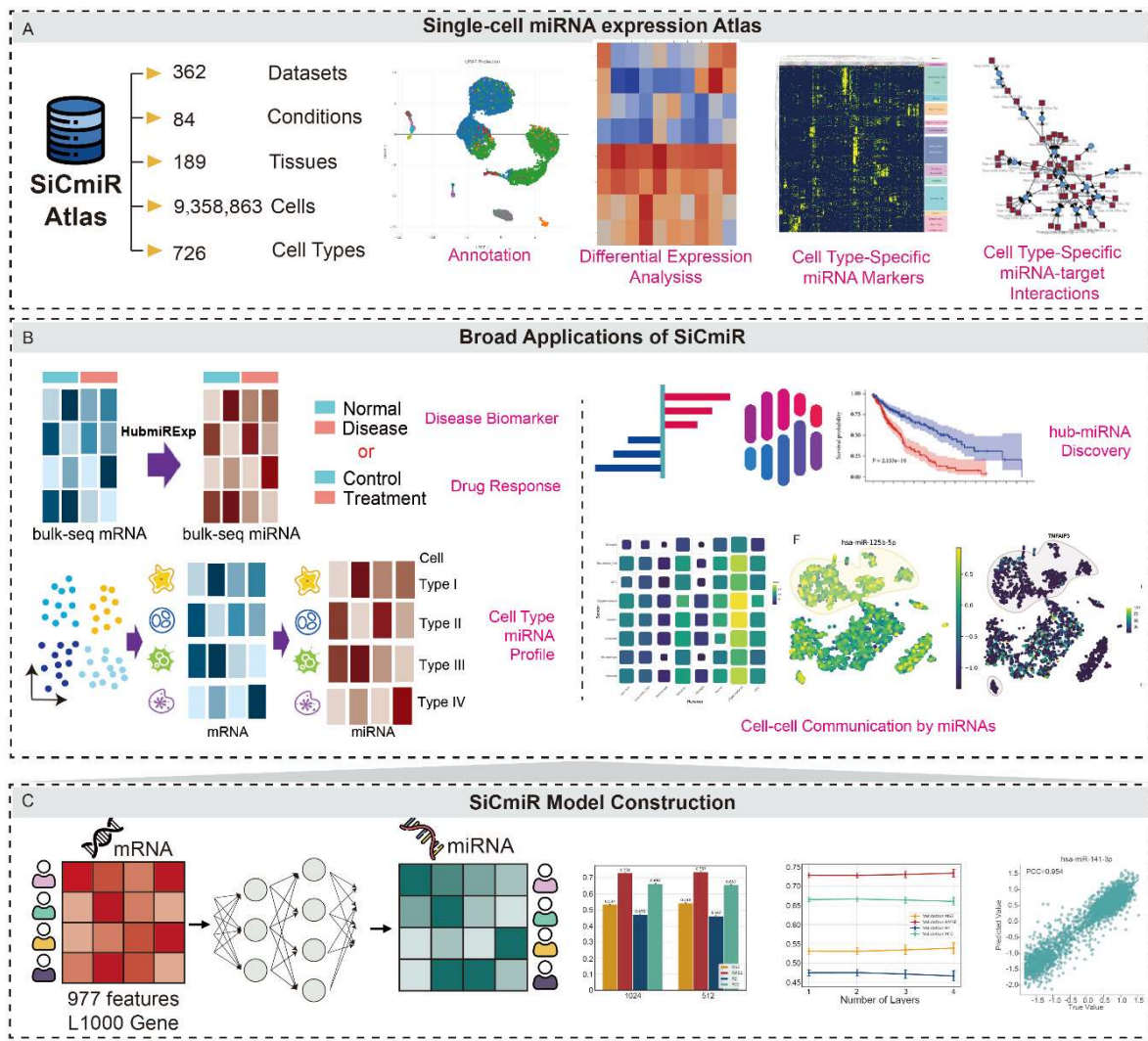

**Figure 1** Schematic representation of the SiCmiR workflow, application and visualization. **A** Construction of SiCmiR Atals storing inferred miRNA expression across various cell types, disease contexts, and four applicable functional modules. **B** Applications in multi-tasks including disease biomarker discovery, drug response prediction, cell-type specific miRNA expression profile prediction, hub-miRNA discovery, and cell-cell communication by miRNA showed robustness of **C** SiCmiR model: leveraging a batch-normalized DNN to infer miRNA profiles from landmark mRNA features.

## Results

### Performance Analysis and Feature Selection of SiCmiR

To evaluate the predictive performance of SiCmiR, we benchmarked multiple model architectures and validated the effectiveness of using 977 input features. Specifically, SiCmiR employs 977 landmark genes from the LINCS L1000 project (see Methods), which are known for their responsiveness to chemical and genetic perturbations, high reproducibility across RNA-seq datasets, and their capacity to infer approximately 81% of non-measured transcript expression levels.[21] The use of these landmark genes also alleviates data sparsity in single-cell applications. We first compared the baseline performance of neural networks, ResNet, and Transformer models with default parameters. Among them, the neural network achieved superior performance (**Table 1**), and was selected for subsequent optimization. Hyperparameters including hidden layer size (1024 nodes), number of layers (2), dropout rate (0.3), and learning rate (0.4) were systematically tuned (**Figure 2A–D**). To assess feature importance, models were trained using different gene sets: the 977 L1000 genes, and the top 1,000, 5,000, 10,000, and all (n = 20,062) variable protein-coding genes. Using L1000 features, the mean PCC across all miRNAs in the training and test sets reached 0.75 and 0.67, respectively (**Figure 2E–F**). Three-fold cross-validation confirmed the robustness of this performance, with average miRNA PCCs of 0.75 ± 0.00067 (training) and 0.67 ± 0.00073 (test), and sample-level PCCs of 0.72 ± 0.09583 (training) and 0.63 ± 0.13707 (test) (**Figure S1**).In comparison across feature sets, the average miRNA PCCs in the training set were 0.79 (L1000), 0.79 (top 1,000), 0.81 (top 5,000), 0.82 (top 10,000), and 0.82 (all genes). Corresponding test set PCCs were 0.67, 0.66, 0.67, 0.67, and 0.68. Notably, when the number of features was limited to ~1,000, the L1000 landmark genes consistently outperformed variably selected gene sets, while offering substantial reductions in computational cost.

**Table 1 Neural network model outperformed among basic models.**

|  | MSE | RMSE | $R^2$ | PCC |
|---|---|---|---|---|
| **Neural Network** | **0.522** | **0.722** | **0.484** | **0.673** |
| ResNet | 0.648 | 0.805 | 0.359 | 0.578 |
| Transformer | 0.863 | 0.929 | 0.146 | 0.359 |

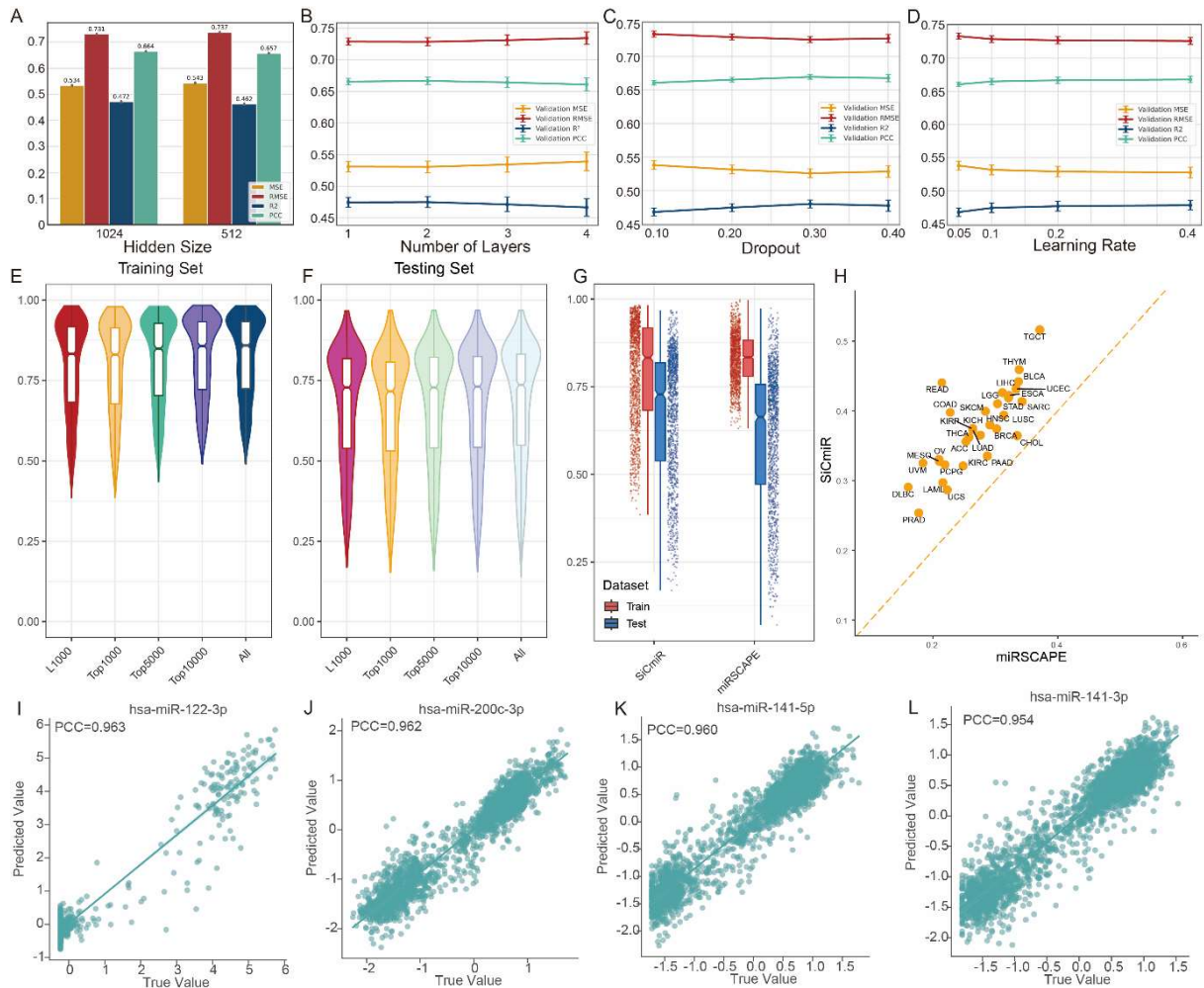

**Figure 2** Performance evaluation on datasets. Mean square error (MSE), root mean square error (RMSE), Pearson correlation coefficient (PCC), and R-square was used for model performance evaluation. **A-D** compares the performance of models with number of nodes in hidden layers to be 512 and 1024, number of hidden layers ranging from 1-4, dropout rate ranging from 0.1 to 0.4 in SGD optimizer, and learning rate ranging from 0.05 to 0.4. **E-F** Average PCC of each miRNA in training set and test set using mRNA features of L1000 landmark genes, top 1000, 5000, 10000 variable mRNAs and all mRNAs. **G-H** Performance of SiCmiR excels miRSCAPE overall and across all cancer types. **I-L** High correlation between observed and predicted values in SiCmiR, highlighting miRNAs with the top performance.

SiCmiR also achieved higher predictive accuracy than miRSCAPE [19] on an independent test set, with a PCC of 0.67 versus 0.61 (**Figure 2G**). Importantly, the inference time of SiCmiR using the pre-trained model was only 2.23 seconds, in stark contrast to over 2 hours required by miRSCAPE due to its on-the-fly model training. To ensure broad applicability, a model trained on 6,462 TCGA pan-cancer samples was employed for all downstream analyses. This pan-cancer model consistently outperformed cancer-specific models and miRSCAPE across diverse cancer types (**Figure 2H** and **Figure S2**). While models trained on individual cancer types were also constructed to account for potential context-specific miRNA–mRNA regulation (**Figure S3**), the pan-cancer model demonstrated superior generalizability and was thus selected. For single-miRNA performance, the highest predictive PCC was 0.984 for hsa-miR-21-5p. Examples of true versus predicted expression levels for four top-ranked miRNAs are shown in **Figure 2I–L**.

**Cross-Cancer Generalization, Bulk-seq and scRNA-seq Case Studies**

To validate the applicability of our method to single-cell miRNA expression prediction, we first analyzed immortal cell line data, assuming homogeneous expression across cells such that bulk-seq profiles represent average single-cell expression. SiCmiR successfully predicted biomarker expression for K562, 293T, HeLa, and A549 cells (**Figure 3A-D**). We then applied SiCmiR to scRNA-seq data from pancreatic ductal adenocarcinoma (PDAC) provided by Peng et al.,[26] which includes 57,530 cells--41,986 from 24 PDAC samples and 15,544 from 11 normal tissues. Cell types were annotated using marker genes from the original study (**Figure 3E** and **Figure S4A**). Following the benchmarking strategy of Olgun et al.,[19] we used a list of 101 dysregulated miRNAs from Mazza et al.,[27] of which 90 are included in our dataset. As type 1 ductal cells (DC1) is reported to be less malignant than type 2 ductal cells (DC2),[26] miRNA expression of Type 2 ductal cells (DC2) in tumours is compared to those of Type 1 ductal cells (DC1) and of their precursor type of cells, acinar cells. SiCmiR correctly predicted 66 miRNAs (sensitivity: 0.73) using pooled data and 28 miRNAs (0.31) using single-cell data directly in DC2 cells compared to DC1 or acinar cells. For the 39 dysregulated miRNAs also covered by miRSCAPE, SiCmiR correctly predicted 29/37 (0.78), while miRSCAPE predicted 29/39 (0.74) (**Supplementary File 1)**. Overall, SiCmiR achieved 0.73 sensitivity (66/90) and 0.65 accuracy (66/101), whereas miRSCAPE achieved 0.29 (29/101). For example, hsa-miR-30b-3p was highly expressed in DC1 compared to DC2 (**Figure 3F**), consistent with higher expression in normal than tumor tissues in bulk-seq. hsa-miR-21-5p was overexpressed in DC2 and MUC5$^+$ DC1 than MUC5$^-$ DC1, in agreement with known PDAC progression **(Figure 3G)**. In all, the assessment of model on PDAC data shows that our model enables detecting miRNAs as biomarkers to indicate cell types and cell states difference for single-cell sequencing data and reaching state-of-the-art.

To further test generalizability for cancer types that are not included in the TCGA data, we applied SiCmiR to ACTH-secreting pituitary adenoma (PA) scRNA-seq data from Zhang et al.,[28] selecting two high-quality samples (SRR13973073, SRR13973076). T-SNE and marker-based annotation by marker genes from original study identified 3,314 PA cells and 1,948 stromal cells (**Figure 3H** and **Figure S4B**). Using stromal cells as baseline, SiCmiR predicted 55/75 reported dysregulated miRNAs[29-32] with sensitivity 0.73 using pooled data and |log2FC| ≥ 0.25. When raising the threshold to |log2FC| ≥ 1, sensitivity dropped to 0.53. The predicted expression of hsa-miR-136-3p and has-miR-410-3p (Figure 3I-J) was highly expressed in stromal cells rather than PA cells as expected.[33] Using un-pooled single-cell data, 46 miRNAs were predicted (0.61 sensitivity); among these, 34 matched the expected trend (0.69) (**Supplementary File 2**). Among top 414 miRNAs by prediction PCC ⩾ 0.8, 49 were dysregulated and 30/48 were predicted using cell-type averages (0.625).These results show SiCmiR's applicability to cancer types not present in TCGA training data., which indicates that the miRNA expression pattern regulated by mRNAs are learnt by our model and the application of our model are not only restricted to cancer types of TCGA but cancer types in variety.

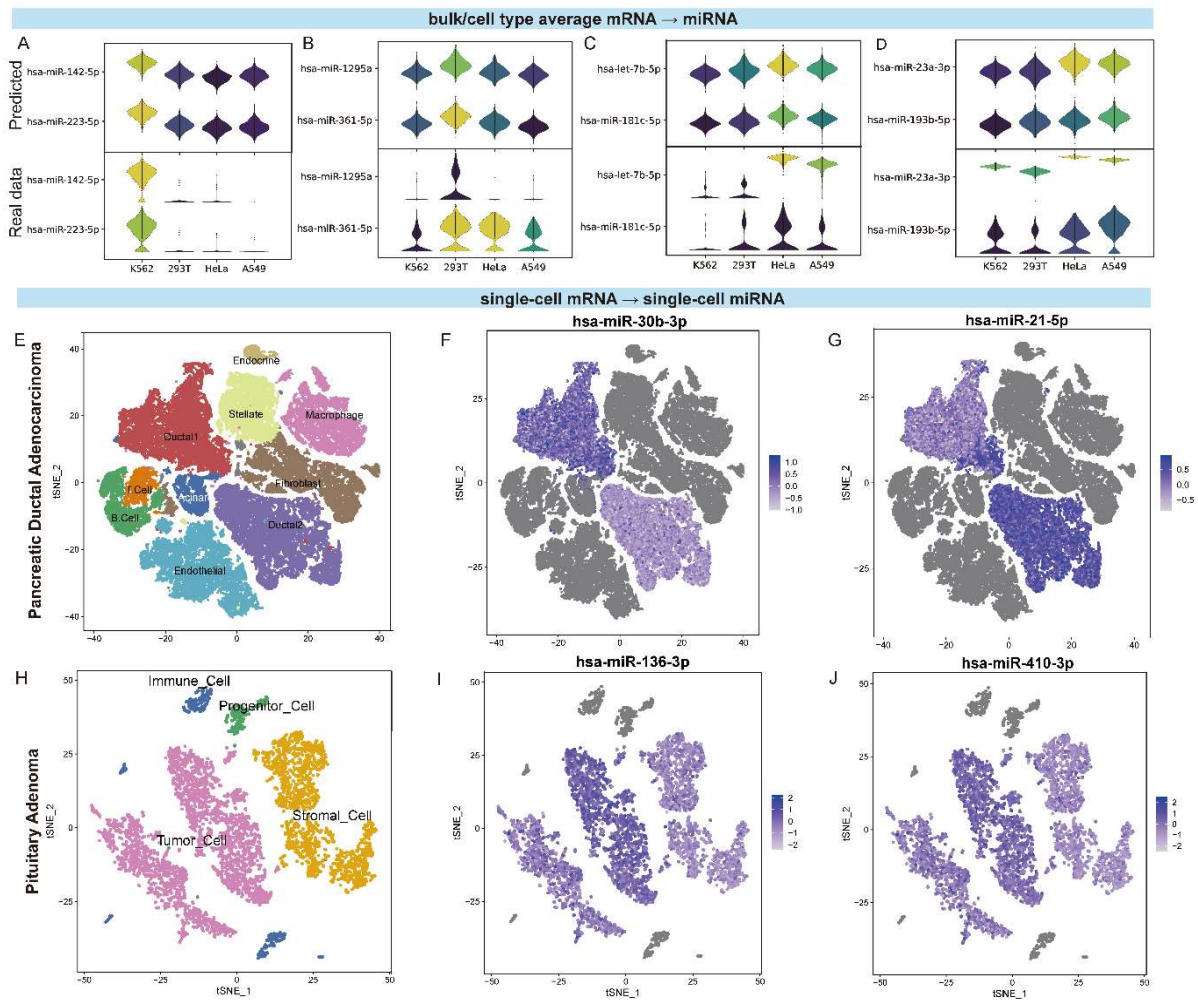

**Figure 3** Expression prediction of miRNAs based on cell line RNA-seq and scRNA-seq data consistent with the literature evidence on real datasets. **A-D** miRNA biomarkers of different cell line inferred based on bulk RNA-seq data (upper) have the same expression pattern as in bulk miRNA-seq data (lower). **E** T-SNE shows the annotation of cells in PDAC scRNA-seq dataset. **F-G** Feature plot of inferred expression profile of hsa-miR-30b-3p and hsa-miR-21-5p and shown significant difference in DC1, MUC5$^+$ DC1, and malignant DC2. **H** T-SNE shows the annotation of cells in PA scRNA-seq dataset. **I-J** Feature plot of inferred expression profile of hsa-miR-136-3p and has-miR-410-3p showed significant difference between tumor cells and stromal cells as in real data.

SiCmiR also applies to bulk tissue and drug-perturbed samples. For hepatocellular carcinoma (HCC) vs. normal tissue in African American (AA) (n=8 per group), 19 of 24 DEmiRs from Varghese et al.[34] were recovered (sensitivity 0.79), with 9 significant ($p < 0.05$). Of 13 DEmiRs with PCC ≥ 0.8 in the test set, 12 were correctly predicted (sensitivity 0.92), including hsa-miR-139-3p/5p and hsa-miR-378d (**Figure S5A**). These results demonstrate SiCmiR's effectiveness in identifying DEmiRs in cancer tissue. We further applied SiCmiR to bulk RNA-seq of A549 cells treated with *Cinnamomi Ramulus* (a kind of traditional Chinese medicine) at 3 concentrations (n=2 each) generated by our lab with DEmiRs valicated by qPCR.[35] Seven DEmiRs were identified, including five (e.g., hsa-miR-25-3p, hsa-miR-183-5p, hsa-miR-218-5p, hsa-miR-27a-3p, hsa-miR-24-3p) with PCC ≥ 0.8 in the test set. hsa-miR-218-5p and hsa-miR-576-5p were significantly downregulated, while hsa-miR-27a-3p and hsa-miR-24-3p showed decreasing trends with CR concentration (**Figure S5B** and **Supplementary File 3**). These results demonstrate SiCmiR's effectiveness in identifying drug-perturbed DEmiRs.

**SiCmiR Atlas Construction and Software Implementation**

To demonstrate the utility of our method, we constructed the SiCmiR Atlas, which integrates 9.36 million single cells from 362 publicly available scRNA-seq datasets spanning 189 anatomically distinct human tissues across 26 major organs as defined by Cell Ontology[22] (**Figure 1A** and **Figure 4A**). Harmonized cell-type annotated cells according to origin studies, yielding 726 unique cell identities from deeply embedded tissue-specific sub-types to broadly shared immune lineages. Clinical metadata were grouped into 84 physiological or disease conditions distributed over 12 broad disease categories. Based on this comprehensive resource, we implemented four fully integrated analysis modules: (i) Data integration and annotation – storing harmonized raw count matrices and cell-type labels; (ii) Biomarkers identification – provides interactive summaries of lineage representation across tissues and conditions; (iii) miRNA/mRNA visualization and differential analysis – supporting rapid visualization, expression comparison, and biomarker discovery for cell-type-enriched or disease-associated miRNAs; (iv) In-built MTI network builder infers miRNA–target interaction (MTI) graphs by integrating target-site predictions by TargetScan,[23] miRWalk,[24] miRDB,[25] and experimentally validated evidence from miRTarBase. In particular, the differential analysis module supports contrastive analysis between disease and normal tissues, allowing users to identify condition-specific dysregulation of miRNAs at single-cell resolution across distinct cell types and disease contexts. Together, these results demonstrate that SiCmiR Atlas delivers a harmonized and annotated database of single-cell miRNA biology. It supports interactive querying (**Figure 4B**), visualizes cell cluster distributions via UMAP (**Figure 4C**) and provides a coherent set of tools for interactive exploration, biomarker discovery and construction of cell-type-resolved regulatory networks (**Figure 4D-E**). Notably, to our knowledge SiCmiR Atlas represents the first publicly available resource dedicated specifically to single-cell mature miRNA expression, providing a scalable, data-driven foundation for both mechanistic studies and translational applications including diagnostic biomarker development and therapeutic target prioritization. Future releases will incorporate additional datasets including spatial transcriptomics and allow user-submitted data for annotation and comparative analysis.

To further demonstrate the analytical power of SiCmiR Atlas, we systematically identified cell type-specific miRNA biomarkers across multiple tissues and conditions. By aggregating predicted miRNA expression profiles from 726 annotated cell types, we prioritized miRNAs that exhibited consistently high expression within specific lineages—such as epithelial cells, endothelial cells, fibroblasts, oligodendrocytes, B cells, T cells/natural killer cells, myofibroblasts, neurons and myeloid compartments—while remaining low in unrelated cell types. A representative heatmap (**Figure 4E**) highlights a panel of these cell-type-enriched miRNAs, revealing robust and recurrent expression patterns across diverse biological contexts. This analysis not only confirms known markers (e.g., miR-126–5p in endothelial cells, miR-141-3p in smooth muscle cells), but also uncovers novel candidates with potential roles in cell identity and function. These conserved signatures offer a valuable reference for miRNA-based cell-type annotation, facilitate deconvolution of bulk miRNA data, and may serve as entry points for studying regulatory circuits and therapeutic targeting in specific cellular compartments.

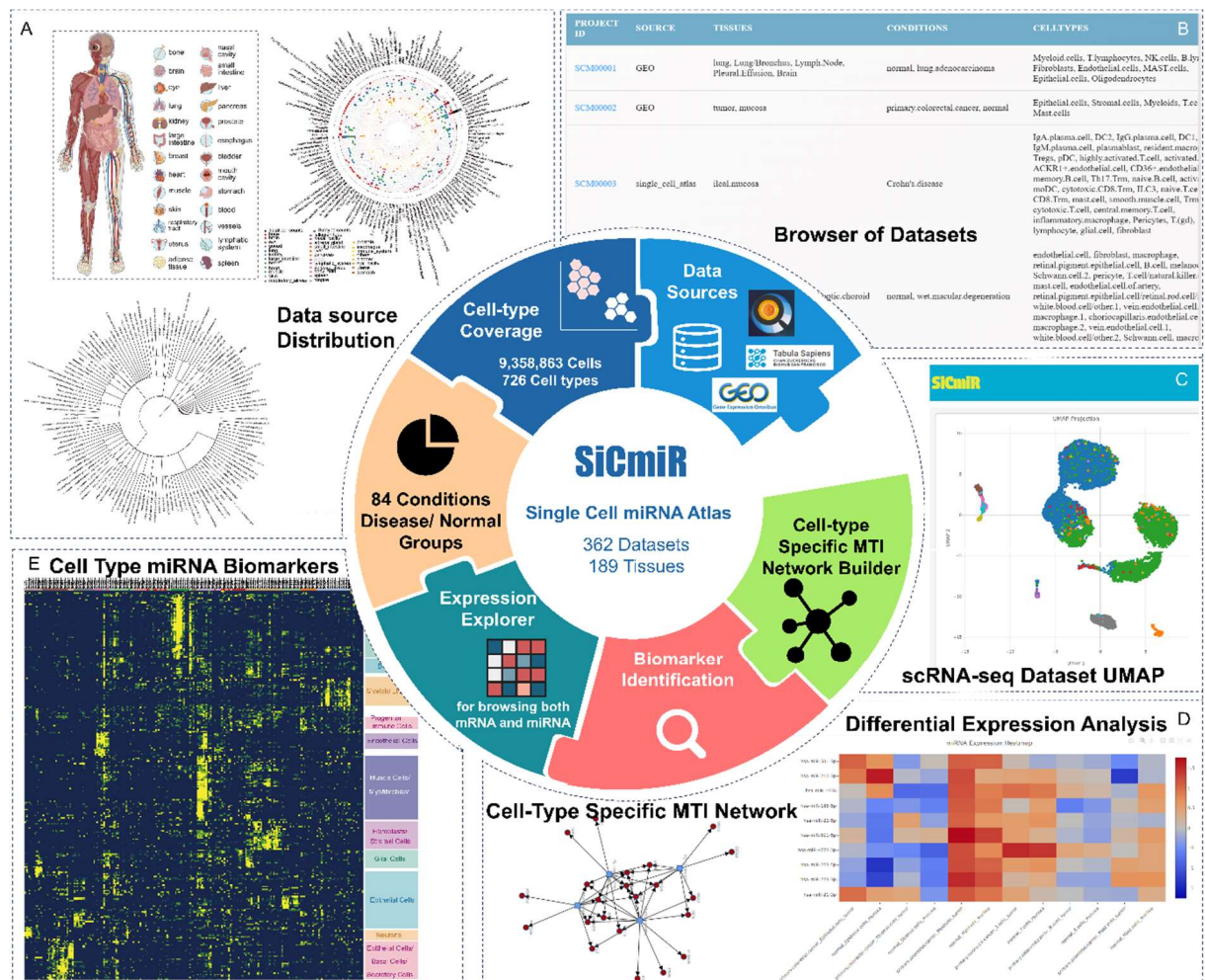

**Figure 4** Architecture, data coverage and core analysis modules of the SiCmiR atlas. **A** Overview of tissue type, cell type distribution and hierarchical clustering dendrogram of cell population of data in SiCmiR Atlas. **B** Example of browser of datasets. **C** Example of T-SNE demonstrating cell type distribution of each dataset. **D** Function of differential expression analysis for miRNA and mRNAs in each dataset to identify cell-type specific biomarkers and to construct cell-type specific MTI network interactively in the webpage. **E** Identified common cell type miRNA biomarkers by SiCmiR Atlas.

**SiCmiR Discovered Hub-miRNAs as Cancer Biomarkers**

As the higher the correlation between miRNA and mRNA expression, the better the pattern of expression can be extracted, and the more active and tight regulation between pairs of them.

We therefore defined miRNAs with a Pearson correlation coefficient (PCC) ≥ 0.80 as hub miRNAs for downstream enrichment and conducted Shapley Additive exPlanations (SHAP) analysis to elucidate their role as prognostic biomarkers. In the independent test set, 414 miRNAs met this threshold, displaying reproducible expression profiles across 33 cancer types and implying tight regulation by their target mRNAs. Among these 414 miRNAs, 105 mature pairs (210 mature miRNAs) originate from the same pre-miRNAs (e.g., hsa-miR-141-3p/5p), and many belong to the same families or primary transcripts (e.g., the miR-200 and miR-302 families). Compared with miRNAs of PCC < 0.8, these hub miRNAs form denser cancer-associated networks: their mean degree is 11.12, ~2.4-fold higher than that of other miRNAs (4.63; **Figure 5A**). Gene Ontology enrichment of their targets highlights pathways central to oncogenesis, progression, and metastasis (**Figure 5B**).

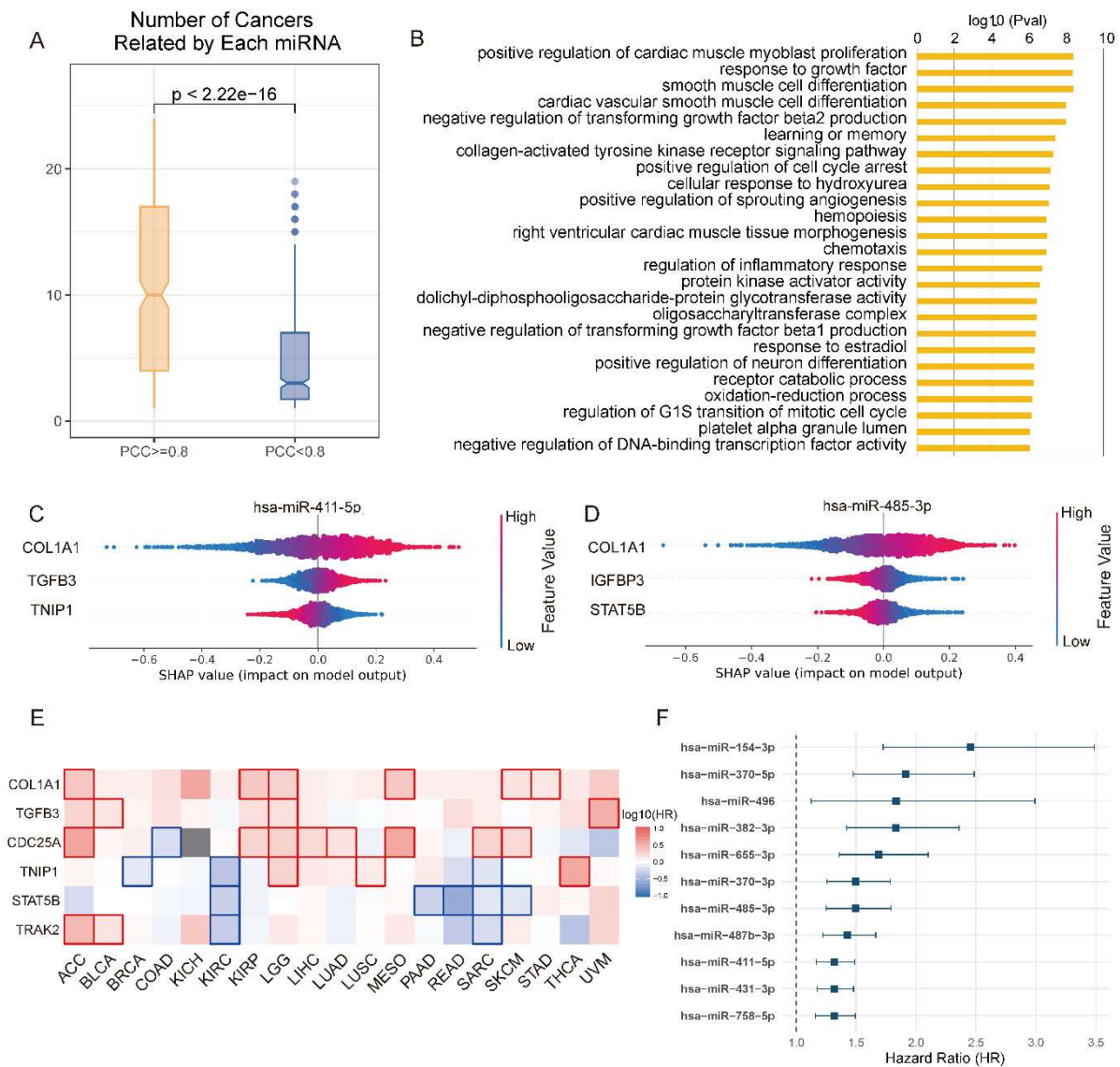

**Figure 5** Identify hub-miRNAs by SiCmiR. **A** Heatmap of miRNA expression in liver cancer vs. normal tissues. Left: Sequencing results; Right: Prediction results. **B** Heatmap of relative miRNA expression in TCM-treated vs. untreated cell lines. Left: Sequencing results; Right: Prediction results. **C** Number of related types of cancers (degree in network) of each miRNA. miRNA with no reported related cancers were excluded, concerning miRNA PCC>=0.8 (n=385) and miRNA PCC < 0.8 (n=356). Single-end Wilcoxon ranked-sum test was used to calculate the significance of difference between miRNAs with PCC>=0.8 (n=385) or PCC < 0.8 (n=528). **D** Top 25 GO enrichment analysis for miRNAs with PCC >=0.8. **E-F** The contribution of landmark genes as features to expression of miRNAs hsa-miR-411-5p and hsa-miR-485-3p. **G** Survival analysis of contributing features secondary to COL1A1 for their association with survival in cancers. **H** Survival analysis of miRNAs in KIRC.

To interpret model predictions, we applied SHAP to quantify each feature's contribution[36] (**Figure S6A**). Network analysis of SHAP-weighted edges revealed 12 functional modules (**Figure S6B**; **Supplementary File 4**). In module 10—driven by features *COL1A1*, *CDC25A*, and *GLI2*—miRNAs cluster on chromosome 14 (**Figure S6C**); 32 of 41 COL1A1-contributing miRNAs (e.g., hsa-miR-127-3p/5p, hsa-miR-134-5p, hsa-miR-136-3p/5p) reside here. Target enrichment links this module to metastasis-related processes, including angiogenesis, extracellular-matrix remodeling, and epithelial–mesenchymal transition (**Figure S6D**). *COL1A1* shows a strong, positive, and significantly larger SHAP contribution than any secondary feature for these 32 miRNAs (**Figure 5C-D**; **Figure S7**). Survival analysis demonstrates that *COL1A1*, *TGFB3*, *CDC25A*, *TNIP1*, *STAT5B*, and *TRAK2*—the top contributors for these miRNAs—correlate with prognosis in kidney renal clear-cell carcinoma (KIRC) and kidney renal papillary carcinoma (KIRP) (**Figure 5E**). High expression of the associated miRNA set predicts markedly poorer survival in KIRP (**Figure 5F**). Given that renal-cell carcinoma progression depends on angiogenesis, invasion, and migration, these findings align with the pathways enriched for COL1A1-linked miRNAs and collectively illustrate how model interpretation uncovers hub-miRNA/mRNA axes driving cancer development.

**SiCmiR Unlocks EV-Mediated Communication Maps in Glioblastoma**

miRTalk[37] delineates how extracellular vesicle (EV) small RNA cargo remodels the tumour niche by coupling a sender-miRNA "secreting score" with receiver cell RISC activity inferred from mRNA data. However, mature miRNA abundance is often uncoupled from miRNA gene expression (**Figure S8**), limiting this approach. To overcome this constraint, we integrated SiCmiR inferred single cell miRNA profiles into the miRTalk framework and re-evaluated cell to cell communication in glioblastoma (GBM). After quality control, 3,497 cells were embedded by t-SNE (**Figure 6A**), resolving eight canonical lineages, including malignant cells,

OPC-like cells and their brain-resident stromal counterparts. Summing significant edges produced a sender receiver matrix (**Figure 6B**) that highlighted pronounced traffic from malignant cells and macrophages, whereas oligodendrocyte progenitor-like cells (OPCs) and neurons acted mainly as sinks. We retained 114,501 high confidence miRNA–target pairs ($P <$ 0.05; **Supplementary File 5**). Incorporating SiCmiR-inferred mature-miRNA abundance into the miRTalk workflow markedly expands both the breadth and biological coherence of the predicted EV-mediated miRNA–target network (**Table 2**). This SiCmiR-enhanced workflow yielded 114,501 high-confidence miRNA–target interactions—>20-fold more than the original proxy analysis—and tripled the chance that an edge displayed the expected negative miRNA–mRNA correlation (36.9 % vs 15.6 %; Fisher's OR = 3.17, 95 % CI 2.94–3.42, $P < 2 \times 10^{-16}$). The average interaction score climbed nearly 50-fold (0.04494 vs 0.00095), reflecting both a denser and more confident interaction landscape, and the aggregate repression effect strengthened slightly (Cliff's δ -0.21; one-sided Wilcoxon $P \approx 0.011$). A heat-map of sender scores ≥ 0.07 for individual miRNAs (Figure 6C) reveals marked cell-type heterogeneity. At single-miRNA resolution, SiCmiR pinpoints lineage-restricted EV cargoes that orchestrate tumour ecology. hsa-miR-125b-5p emerged as specific, exhibiting an elevation in The Cancer Genome Atlas (TCGA) GBM samples versus lower-grade glioma (LGG) and non-malignant cells (**Figure 6D**). Within TCGA-GBM, miR-125b-5p levels correlated negatively with the expression of its validated targets (ρ = –0.23, P = 0.048; **Figure 6E**), indicating effective target repression in bulk tissue populations. These results were also reported by X Shao et al.[37] Feature overlays confirmed that miR-125b-5p is enriched in sender malignant clusters, while its target TNFAIP3 is reciprocally expressed in malignant clusters themselves and neighbouring astrocytes (**Figure 6F**). These correlations corroborate an autocrine loop wherein miR-125b-rich EVs reinforce lipid metabolism signalling and suppressing the programmed cell death within the tumour core (**Figure 6G**). Likewise, hsa-miR-10b-5p suppresses five pro-apoptotic genes in recipient OPC-like cells (**Figure 6H–J**) which are usually the progenitor of malignant cells in GBM, consistent with previous reports that miR-10b confers survival advantages and invasive phenotypes.[38] Conversely, macrophage-enriched hsa-miR-21-5p exports oncogenic signals to malignant clusters, correlating positively with B3GNT5, ICAM1 and TNFAIP3 (**Figure 6K–L**). Collectively, these gains demonstrate that supplying mature-miRNA expression predicted by SiCmiR not only inflates network coverage but, more importantly, substantially enhances the biological plausibility of miRTalk's intercellular miRNA-target predictions, thereby providing a higher-resolution, functionally consistent view of EV-mediated communication.

**Table 2 SiCmiR Integration Dramatically Broadens and Refines EV-Mediated miRNA–Target Networks.**

| Metric | miRTalk (gene-proxy) | SiCmiR (mature-miRNA) | Fold-change / Gain | Statistical test |
|---|---|---|---|---|
| Total high-confidence edges | 5 390 | 114 501 | × 21.2 | - |
| Negative-correlated MTI ($\rho < 0$, p-value < 0.05) | 840 | 42 273 | × 50.3 | Fisher exact test OR = 3.17 (95 % CI = 2.94 -3.42), $P < 2.2e-16$, |
| Negative-correlated MTI proportion | 15.6 % | 36.9 % | +21.3 pp | Same Fisher test |
| Repression effect size (Cliff's δ on receiver-cell) | -0.2 | -0.21 | Slight increase | One-sided Wilcoxon $P = 0.0128$, 0.0107 |
| Average MTI score (by miRTalk) | 0.00095 | 0.04494 | × 47.23 | One-sided Wilcoxon $P < 2.2e-16$ |

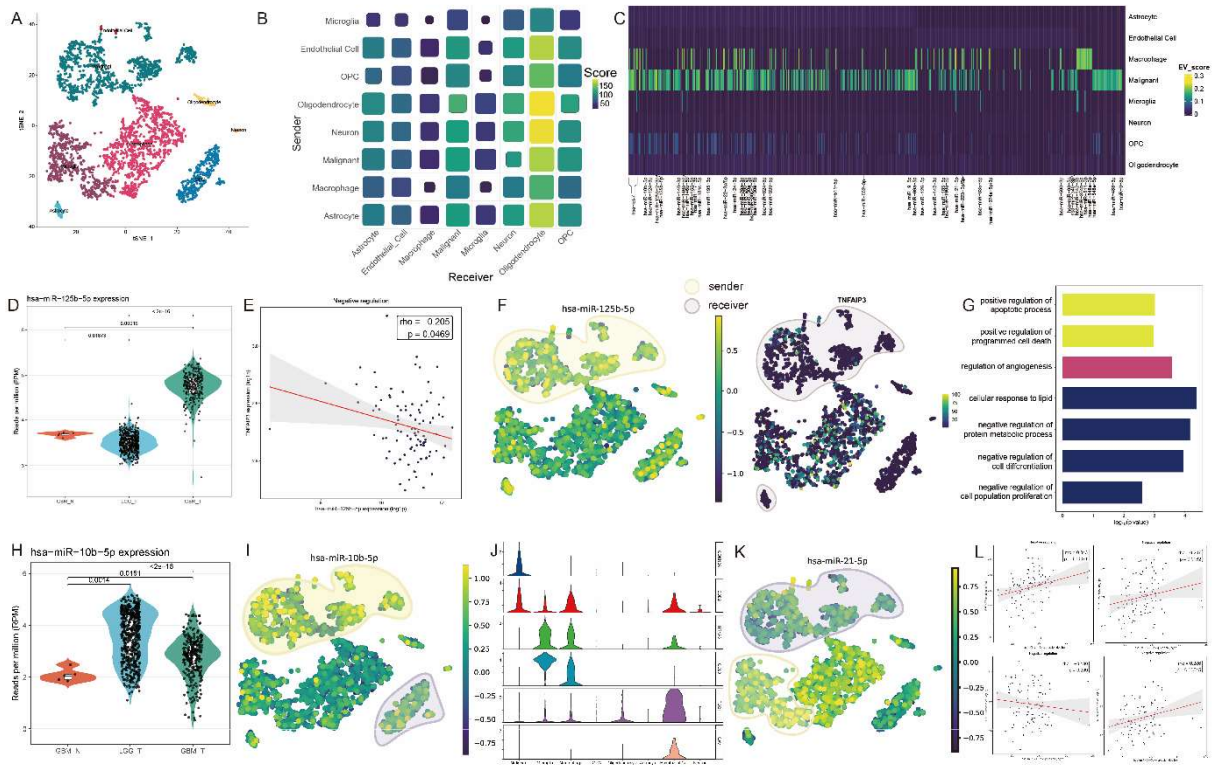

**Figure 6** SiCmiR enable the unraveling of cell-cell communication in GBM via extracellular vesicle-mediated transfer. **A** T-SNE embedding of 3,497 cells clustered into eight lineages. **B** Sender-receiver matrix summarizing cross-talk strength; circle size denotes the number of significant miRNA-target edges and color indicates the cumulative communication score. **C** Heatmap of high-confidence miRNA–target pairs (rows) and ranked by EVmiR score across cell types (columns). **D** Violin plot of hsa-miR-125b-5p expression (log-CPM) in TCGA-GBM, LGG versus all non-malignant cells; p-values by two-tailed Wilcoxon tests. **E** Spearman correlation between hsa-miR-125b-5p levels and the mean EVmiR score of its targets TNFAIP3 ($\rho$ = -0.23, $P$ = 0.048) within TCGA-GBM cohort; shaded band, 95 % CI. **F** Feature TSNEs: left, hsa-miR-125b-5p; right, representative target gene TNFAIP3. Yellow indicates sender cells; purple indicates receiver cells. **G** GO enrichment for genes negatively correlated with hsa-miR-125b-5p (-log10 p-value). **H** Expression landscape of hsa-miR-10b-5p same as **D**. **I** T-SNE feature map for hsa-miR-10b-5p with legend the same as **F**. **J** Expression of five apoptosis-related hallmark as hsa-miR-10b-5p target genes across cell types. **K** T-SNE feature map of hsa-miR-21-5p, with expression confined to a malignant sub-cluster. **L** Scatter plots showing Spearman correlations between hsa-miR-21-5p and 4 canonical targets (*EGFR, MMP2, TGFBI, BTG2*) within TCGA-GBM cohort; red lines indicate fitted regressions, shaded bands the 95 % CI.

## Discussion

This study introduces SiCmiR, an advanced computational framework designed to infers miRNA activity from only 977 landmark genes and scales these predictions into SiCmiR-Atlas, the first open repository of single-cell mature-miRNA landscapes. Three points deserve emphasis. By shrinking the input space from ~20 000 to 977 genes, SiCmiR mitigates zero-inflation and attains state-of-the-art accuracy across 33 TCGA tumor types and multiple scRNA-seq datasets. This addresses a central limitation of reliance on full-transcriptome features leads to performance decay in sparse single-cell matrices. Applying the model to a harmonized compendium of public datasets, we assembled SiCmiR Atlas, which warehouses single-cell miRNA predictions, and cell-type metadata in a user-friendly webpage. Interactive modules enable users to visualize expression patterns, mine hub-miRNAs, gene biomarkers and export direction-resolved miRNA-target networks without local computation. Proof-of-concept vignettes in hepatocellular carcinoma (HCC), glioblastoma (GBM) and ACTH-secreting pituitary adenoma show that atlas-level predictions (i) recover literature-supported oncogenic miRNAs, (ii) reveal candidate hub-regulators with prognostic value, and (iii) illuminate EV-mediated crosstalk among malignant and stromal populations.

In our results, accurate predictions of and the identification of correlation between mature miRNA expression levels within the same family are reasonable, given their common transcription from the same precursor primary miRNA or closely located chromosomal regions and sharing same transcription factors. Importantly, this suggests that our neural network model effectively captures mature miRNA expression patterns without explicitly considering post-transcriptional splicing, modification, or degradation processes.

Case-study shows the utility of SiCmiR identifying miRNAs with potential biological significance in cancers and cell types. It facilitates researchers to understand the cell types that dysregulated miRNAs actually performing functions in during the cancer development and prognosis. In miRNAs that correctly detected by SiCmiR, hsa-miR-21 that involved in reducing apoptosis of tumor cells and promoting their proliferation[39] are up-regulated in all acinar cells, DC1, and DC2. hsa-miRNA-221 and hsa-miRNA-222 that promotes PDAC cancer cell invasion are up-regulated in both DC1 and DC2 in cancer samples. The result shows that by only down-regulating hsa-miR-146a in DC1 can promote the invasion of PDAC cancer. Newly discovered unreported dysregulated miRNAs can also provide further insights into the potential therapeutic targets for cancers. For example, hsa-miR-147b-3p was found to significantly up-regulated in only DC2 but not DC1 or acinar cells. Its targets, including *SDHD*, *NDUFA4*, and *ALDH5A1*, are involved in the cell respiration process.[40] Deficient of *SDHD* in pancreatic ductal cancer were reported to be associated with the accumulation of ROS, leading to abnormalities in tumor cell metabolism.[41] As DC2 are more malignant compared with other cell types,[26] the overexpression of hsa-miR-147b-3p could be one of the causations of malignancy. Findings in case-studies will be essential to confirm by experimental validation. Future improvements should address features beyond the 977-gene panel, such as non-coding RNAs, RNA-binding proteins and epigenetic modifiers, through multi-omic inputs or adaptive feature selection.[42, 43] Enhanced model interpretability may reveal further regulatory motifs. Additionally, EVmiR quantification presently assumes linear additivity and homogeneous vesicle uptake; incorporating spatial context and vesicle proteomics promises finer-grained directionality estimates.

**Conclusion**

SiCmiR bridges a critical gap in single-cell transcriptomics by enabling robust, fine-grained inference of miRNA activity from a compact 977-gene feature set. This design not only mitigates dropout-related noise that hampers transcriptome-wide models but also accelerates computation, facilitating routine incorporation of miRNA layers into single-cell analyses. By scaling the pipeline across diverse publicly available datasets, we generated SiCmiR Atlas, a freely accessible repository that integrates predicted miRNA abundance and cell-type annotations. Proof-of-concept studies in hepatocellular carcinoma, glioblastoma and ACTH-secreting pituitary adenoma demonstrate the resource's capacity to uncover candidate hub-miRNAs and to chart extracellular-vesicle-mediated regulatory circuits with single-cell resolution. Several avenues remain for future refinement. Nevertheless, the present work delivers both a method and a community resource that together lay a foundation for systematically dissecting miRNA-driven cell–cell interactions and accelerating their translational exploitation in precision medicine.

**Materials and methods**

**Data collection for model training, testing and case-study**

Matched bulk Fragments Per Kilobase of transcript per million mapped reads (FPKM) RNA-Seq and reads per million mapped reads normalized (RPM) normalized miRNA-Seq gene expression data for cancers and normal samples were retrieved from TCGA using UCSC Xena at https://xenabrowser.net/ [44] (**Figure S9**). miRNAs were selected from the union of samples with at least one non-zero expressing sample in TCGA data. mRNA expression profiles of 977 mRNAs in 978 L1000 genes were extracted. XBP1 in L1000 landmark genes were excluded due to zero count in all samples. 1298 miRNAs out of 1952 miRNAs were selected to filter out miRNAs with zero counts in all samples. Known experimentally validated miRNA target information was gathered from the miRTarBase.[40] The training data set contains 6,462 samples from 33 types of cancers. The rest of samples from TCGA are used as independent validation sets. For independent validation datasets, there are totally 2,768 samples. The ratio of number of samples for each cancer type is around 3 to 1 as the ratio of number of samples in training dataset and test dataset is close to 3:1.

For case-study, bulk RNA-seq and small RNA-seq data were collected. Bulk RNA-seq data for hepatocellular carcinoma was collected from Varghese et al., GEO accession number: GSE176289.[34] Bulk RNA-seq data for non-small cell lung cancer (NSCLC) A5459 cell line is generated by our lab and published by Li et al.[35] The scRNA-seq expression profile of GBM obtained from GEO with accession GSE64465.[45] The expression of each mRNA and miRNA across cells was normalized by z-score across samples before applying SiCmiR model. The scRNA-seq expression profile for PDAC was collected from Peng et al.[26] at PRJCA001063 from https://ngdc.cncb.ac.cn/. The PA scRNA-seq data is retrieved from GEO with accession SRR13973073, SRR13973076. Cells were filtered by nFeature_RNA $\geq$ 200 and percentage of mitochondrial reads $\leq$ 10%. Gene counts were library-size normalized (CPM × 1e4) and log-transformed.

**Machine learning model for miRNA profile prediction**

We've adopted the neural network architecture that predicts miRNA profiling based on the given mRNA expression levels. Denote the training dataset $\mathcal{D} = \{(x^{(1)}, y^{(1)}), ..., (x^{(N)}, y^{(N)})\}$ with a total of $N$ samples, where the $x^{(n)} \in \mathbb{R}^d$ stands for the $d$-dimensional gene expression vector and $y^{(n)} \in \mathbb{R}^m$ represents for the $m$-dimensional vector of miRNA profiling values for the $n$-th sample. The goal is to utilize $\mathcal{D}$ to learn a neural network-based multi-target repression model $\mathcal{F}_\theta(\cdot)$ parameterized by $\theta$ that maps the input gene expression vector $x$ to the output vector $\hat{y}$ of the miRNA values. A two-layer fully connected network (input = 977, hidden = 1024, output = 1298) was implemented in Pytorch (cuda-11.7). Hyper-parameters were tuned by grid search (**Supplementary File 6**). Early stopping after 20 epochs without validation loss improvement. We considered batch normalization,[46] dropout,[47] and rectified linear unit (ReLU)[48] for each hidden layer to avoid overfitting and improve the prediction performance. A detailed schematic diagram for the structures of the adopted neural network model, and ResNet and Transformer model for comparison is illustrated in **Figure S10** and **Supplementary File 6.** 20% of training set are separated randomly by seed = 42, 52, 62 as validation set. To achieve the predictive performance of the regression tasks, the model utilizes a mean squared error (MSE) loss function $l(\cdot)$ as:

$$l(\mathcal{D}, \theta) = \frac{1}{N} \sum_{n=1}^{N} \left\| y^{(n)} - \hat{y}^{(n)} \right\|$$

where the $\hat{y}^{(n)}$ denotes the output prediction for the $n$-th training sample. The supervised loss encourages the model parameter $\theta$ to update and finally be capable to predict miRNA values from gene expression inputs. We implemented the model training by the stochastic gradient descent optimizer with a 0.4 learning rate. The dropout rates are set as 0.3 for the hidden layers. We performed stratified k-fold CV (k = 3) to avoid data leakage. Feature selection (977-gene landmark) was fixed a priori; and thus, CV was not nested.

**Performance Evaluation**

To characterize the predictive performance of our proposed regression model, we adopted the Pearson correlation coefficient (PCC) to measure the consistency between model prediction and the ground-truth miRNA prediction value. The PCC for a specific miRNA regression is defined as:

$$\text{PCC} = \frac{\sum_{n=1}^{N}(y_k^{(n)} - \overline{y_k})(\hat{y}_k^{(n)} - \overline{\hat{y}_k})}{\sqrt{\sum_{n=1}^{N}(y_k^{(n)} - \overline{y_k})^2 \sum_{n=1}^{N}(\hat{y}_k^{(n)} - \overline{\hat{y}_k})^2}}$$

where the $y_k^{(n)}$ and $\hat{y}_k^{(n)}$ stand for $k$-th miRNA ground-truth value and the prediction result of the $n$-th sample from the dataset. The $\overline{y_k} = \frac{1}{N}\sum_{n=1}^{N} y_k^{(n)}$ and $\overline{\hat{y}_k} = \frac{1}{N}\sum_{n=1}^{N} \hat{y}_k^{(n)}$ are the average of true miRNA profiles and prediction value, respectively. Mean square error (MSE) and root mean square error (RMSE) quantify the average squared and root-squared deviations, with RMSE sharing the same scale as the original data.

$$MSE_k = \frac{1}{N}\sum_{n=1}^{N}\left(\hat{y}_k^{(n)} - y_k^{(n)}\right)^2,$$

$$RMSE_k = \sqrt{MSE_k}.$$

The coefficient of determination R-square ($R^2$) revaluates how much of the variance in the ground-truth expression levels is explained by the model;

$$R_k^2 = 1 - \frac{\sum_{n=1}^{N}(\hat{y}_k^{(n)} - y_k^{(n)})^2}{\sum_{n=1}^{N}(y_k^{(n)} - \bar{y}_k^{(n)})^2}, \quad \bar{y}_k = \frac{1}{N}\sum_{n=1}^{N}y_k^{(n)}.$$

$R^2 = 1$ indicates perfect fit, $R^2 = 0$ means the model performs no better than simply predicting the mean, and $R^2 < 0$ implies worse performance than the mean predictor.

**SHAP analysis attributes the contribution of each mRNA to output**

To attribute the contribution of each input feature to the model output, gradient explainer for SHAP analysis was adopted.[36] The average contribution of each feature to each output miRNA in each paired sample was calculated.

**Annotation of miRNA functions and pathway enrichment analysis**

miREAA2 (https://www.ccb.uni-saarland.de/mieaa2) was used for annotation of miRNAs of their roles in different types of cancers.[49] Enrichment analysis of miRNAs was also conducted. Over-represented mode was chosen for annotation of miRNAs in cancers. The network graphs between miRNAs and cancers and the analysis of the network, e.g. degree of nodes, were plot and calculated by Gephi.[50] MetaCore (@Clarivate Analytics, https://portal.genego.com/) and gene ontology[51] by R package clusterProfiler,[52] and Kyoto Encyclopedia of Genes and Genome (KEGG) Pathway database (https://www.genome.jp/kegg/pathway.html) was used for gene enrichment analysis. Default parameters were used for analysis.

**Survival analysis for discovered hub-miRNA**

Survival analysis for miRNAs in interests in cancers are computed by oncomiR (http://www.oncomir.org/).[53] Survival analysis for mRNA in cancers are computed by GEPIA2 (http://gepia2.cancer-pku.cn/).[54] Univariate Cox analysis was applied. Difference of survival rate between low and high expression with P-value <0.05 was regarded as significant.

**Data Processing and Differential expression analysis**

DESeq2[55] with p-value calculated by Wald test were used for differential expression analysis for bulk-seq/predicted-bulk miRNA expression profile. For predicted single-cell miRNA expression profiles, Seurat V4[56] was used to conduct differential expression analysis with p-value calculated by Wilcoxon Rank Sum test (Wilcox). The threshold of p-value is p-value <0.05.

**scRNA-seq data sampling and pooling for case-study**

In case-study part, scRNA-seq data were sampled by cells in each reported cell types and pooled as pseudo-bulk data for better prediction accuracy. The scRNA-seq data was pooled in order to avoid the sequencing bias and sparsity of scRNA-seq or conducted cell type average pooling. For pooling average, cells in each cell types are randomly sampled not replacing 80% of cells for average pooling in one pooled sample, which is the same as the bootstrapping method in Olgun et al.[19]

**Cell-cell communication imputation**

miRTalk[37] is used for imputations with miRNA host genes expression replaced by predicted miRNA expression profiles by SiCmiR as mentioned above and other parameters remaining default.

**Database implementation and Github usage**

SiCmiR Atals webpage at https://awi.cuhk.edu.cn/~SiCmiR/ was built using apache wicket framework on local high performance computational server running a CentOS Linux system.

The model of SiCmiR is also available on Github at https://github.com/Cristinex/SiCmiR/. For the visualization of predicted miRNA level in different subpopulation, R package Seurat V4[56] was used.

**Key Points**

- **State-of-the-art prediction model** We constructed a two-layer neural network trained on 6,462 TCGA samples predicts the activity of 1,298 miRNAs from 977 L1000 gene inputs with a mean Pearson correlation of 0.67, outperforming existing tools in 33 cancer types and generalizing to unseen cancer types in training sets, perturbations and scRNA-seq data.
- **First single-cell mature miRNA atlas** SiCmiR Atlas is, to our knowledge, the world's first public database dedicated to single-cell mature miRNA expression, integrating 9.36 million cells from 362 datasets that span 189 tissues and 82 physiological or disease conditions constructed by applying SiCmiR model.
- **Comprehensive analytical toolkit** The SiCmiR Atlas platform enables hub-miRNA discovery, cell-type-specific biomarker identification and construction of cell-type-resolved miRNA–target interaction networks.
- **Translational relevance** SiCmiR supports mechanistic studies and accelerates translational efforts such as diagnostic biomarker discovery and therapeutic target identification. It can also be supply to identify extracellular vesicles-driven cell-cell communication mediated by miRNAs.

## Competing interests

The authors have declared no competing interests.

**Authors' contributions**

Conceptualization: Hsien-Da Huang; Methodology, X.-X. Cai, Y.-X. Pang, T.-S. Xu, and T.-Y. Lee; Model construction: X.-X. Cai, Y.-X. Pang, T.-S. Xu, and X. Cao; Model Performance Analysis: X.-X. Cai, J.-S. Liao, Y.-D. Chen and T.-S. Xu; Web server construction: X.-X. Cai, J.-J. Ma, J.-S. Liao, H.-Y. Huang, and Y.-G. Chen; Formal analysis: X.-X. Cai, J.-S. Liao, and Y.-X. Pang; Investigation: X.-X. Cai, H.-Y. Huang and H.-D. Huang; Data curation: X.-X. Cai, J.-S. Liao, J.-J. Ma, Y.-D. Chen, T.-S. Xu and Y.-C. Zhang; Writing—original draft preparation: X.-X. Cai and Y.-X. Pang; Writing—review and editing: X.-X. Cai, H.-D. Huang, H.-Y. Huang, T.-Y. Lee and Y.-C.-D. Lin; Visualization: X.-X. Cai, H.-Y. Huang, Y.-X. Pang and Y.-D Chen; Supervision: H.-D. Huang and H.-Y. Huang; Project administration: H.-D. Huang and H.-Y. Huang; Funding acquisition: H.-D. Huang, H.-Y. Huang and Y.-C.-D. Lin.

**Fundings**

This research was funded by Shenzhen Science and Technology Innovation Program [JCYJ20220530143615035]; Guangdong S&T programme [2024A0505050001, 2024A0505050002]; Warshel Institute for Computational Biology funding from Shenzhen City and Longgang District [LGKCSDPT2024001]; Shenzhen-Hong Kong Cooperation Zone for Technology and Innovation [HZQB-KCZYB-2020056, P2-2022-HDH-001-A]; Guangdong Young Scholar Development Fund of Shenzhen Ganghong Group Co., Ltd. [2021E0005, 2022E0035, 2023E0012].

**Acknowledgements**

This research benefited significantly from the interdisciplinary research environment and cutting-edge instrumentation maintained by the Computational Platform of Warshel Institute for Computational Biology. Their ongoing commitment to research infrastructure development has been crucial to our scientific endeavors. Authors are also grateful to the library of The Chinese University of Hong Kong, Shenzhen for providing effective database service.

**Data availability**

The authors declare that data supporting the findings of this study are available within the article and supplementary information. Package for usage of SiCmiR can be retrieved at https://github.com/Cristinex/SiCmiR. Online application is deposited at https://awi.cuhk.edu.cn/~SiCmiR/. Data supporting this study are openly available in Zenodo repository at https://doi.org/10.5281/zenodo.16025109.

# References

1. O'Brien J, Hayder H, Zayed Y, Peng C: **Overview of microRNA biogenesis, mechanisms of actions, and circulation**. *Frontiers in Endocrinology* 2018, **9**(AUG):1-12.
2. Lee LJ, Papadopoli D, Jewer M, del Rincon S, Topisirovic I, Lawrence MG, Postovit L-M: **Cancer Plasticity: The Role of mRNA Translation**. *Trends in Cancer* 2021, **7**(2):134-145.
3. Peng Y, Croce CM: **The role of MicroRNAs in human cancer**. *Signal Transduct Target Ther* 2016, **1**:15004.
4. Kozomara A, Birgaoanu M, Griffiths-Jones S: **miRBase: from microRNA sequences to function**. *Nucleic Acids Res* 2019, **47**(D1):D155-D162.
5. Benesova S, Kubista M, Valihrach L: **Small RNA-Sequencing: Approaches and Considerations for miRNA Analysis**. *Diagnostics (Basel)* 2021, **11**(6).
6. Zhang B, Horvath S: **A general framework for weighted gene co-expression network analysis**. *Stat Appl Genet Mol Biol* 2005, **4**:Article17.
7. Na Y-J, Kim JH: **Understanding cooperativity of microRNAs via microRNA association networks**. *Bmc Genomics* 2013, **14**(Suppl 5):S17.
8. Li C, Dou P, Wang T, Lu X, Xu G, Lin X: **Defining disease-related modules based on weighted miRNA synergistic network**. *Computers in Biology and Medicine* 2023, **152**:106382.
9. Shao T, Wang G, Chen H, Xie Y, Jin X, Bai J, Xu J, Li X, Huang J, Jin Y *et al*: **Survey of miRNA-miRNA cooperative regulation principles across cancer types**. *Briefings in Bioinformatics* 2019, **20**(5):1621-1638.
10. Xu J, Shao T, Ding N, Li Y, Li X: **miRNA-miRNA crosstalk: from genomics to phenomics**. *Brief Bioinform* 2017, **18**(6):1002-1011.
11. Su B, Wang W, Lin X, Liu S, Huang X: **Identifying the potential miRNA biomarkers based on**

multi-view networks and reinforcement learning for diseases. *Briefings in Bioinformatics* 2024, **25**(1):bbad427.

12. Wang S, Li Y, Zhang Y, Pang S, Qiao S, Zhang Y, Wang F: **Generative Adversarial Matrix Completion Network based on Multi-Source Data Fusion for miRNA–Disease Associations Prediction**. *Briefings in Bioinformatics* 2023, **24**(5):bbad270.
13. Alfardus H, de los Angeles Estevez-Cebrero M, Rowlinson J, Aboalmaaly A, Lourdusamy A, Abdelrazig S, Ortori C, Grundy R, Kim D-H, McIntyre A *et al*: **Intratumour heterogeneity in microRNAs expression regulates glioblastoma metabolism**. *Scientific Reports* 2021, **11**(1):15908.
14. Wang N, Zheng J, Chen Z, Liu Y, Dura B, Kwak M, Xavier-Ferrucio J, Lu YC, Zhang M, Roden C *et al*: **Single-cell microRNA-mRNA co-sequencing reveals non-genetic heterogeneity and mechanisms of microRNA regulation**. *Nature Communications* 2019, **10**(1):1-12.
15. Faridani OR, Abdullayev I, Hagemann-Jensen M, Schell JP, Lanner F, Sandberg R: **Single-cell sequencing of the small-RNA transcriptome**. *Nat Biotechnol* 2016, **34**(12):1264-1266.
16. Ji J, Anwar M, Petretto E, Emanueli C, Srivastava PK: **PPMS: A framework to Profile Primary MicroRNAs from Single-cell RNA-sequencing datasets**. *Brief Bioinform* 2022, **23**(6).
17. Engel A, Rishik S, Hirsch P, Keller V, Fehlmann T, Kern F, Keller A: **SingmiR: a single-cell miRNA alignment and analysis tool**. *Nucleic Acids Research* 2024, **52**(W1):W374–W380.
18. Hücker SM, Fehlmann T, Werno C, Weidele K, Lüke F, Schlenska-Lange A, Klein CA, Keller A, Kirsch S: **Single-cell microRNA sequencing method comparison and application to cell lines and circulating lung tumor cells**. *Nature Communications* 2021, **12**(1):1-13.
19. Olgun G, Gopalan V, Hannenhalli S: **miRSCAPE - inferring miRNA expression from scRNA-seq data**. *iScience* 2022, **25**(9):104962.
20. Herbst E, Mandel-Gutfreund Y, Yakhini Z, Biran H: **Inferring single-cell and spatial microRNA activity from transcriptomics data**. *Communications Biology* 2025, **8**(1):87.
21. Subramanian A, Narayan R, Corsello SM, Peck DD, Natoli TE, Lu X, Gould J, Davis JF, Tubelli AA, Asiedu JK *et al*: **A Next Generation Connectivity Map: L1000 Platform and the First 1,000,000 Profiles**. *Cell* 2017, **171**(6):1437-1452.e1417.
22. Diehl AD, Meehan TF, Bradford YM, Brush MH, Dahdul WM, Dougall DS, He Y, Osumi-Sutherland D, Ruttenberg A, Sarntivijai S *et al*: **The Cell Ontology 2016: enhanced content, modularization, and ontology interoperability**. *J Biomed Semantics* 2016, **7**(1):44.
23. McGeary SE, Lin KS, Shi CY, Pham TM, Bisaria N, Kelley GM, Bartel DP: **The biochemical basis of microRNA targeting efficacy**. *Science* 2019, **366**(6472).
24. Sticht C, De La Torre C, Parveen A, Gretz N: **miRWalk: An online resource for prediction of microRNA binding sites**. *PLOS ONE* 2018, **13**(10):e0206239.
25. Chen Y, Wang X: **miRDB: an online database for prediction of functional microRNA targets**. *Nucleic Acids Research* 2020, **48**(D1):D127-D131.
26. Peng J, Sun B-F, Chen C-Y, Zhou J-Y, Chen Y-S, Chen H, Liu L, Huang D, Jiang J, Cui G-S *et al*: **Single-cell RNA-seq highlights intra-tumoral heterogeneity and malignant progression in pancreatic ductal adenocarcinoma**. *Cell Research* 2019, **29**(9):725-738.
27. Mazza T, Copetti M, Capocefalo D, Fusilli C, Biagini T, Carella M, De Bonis A, Mastrodonato N, Piepoli A, Pazienza V *et al*: **MicroRNA co-expression networks exhibit increased complexity in pancreatic ductal compared to Vater's papilla adenocarcinoma. *Oncotarget* 2017, **8**(62):105320-105339.

28. Zhang D, Hugo W, Redublo P, Miao H, Bergsneider M, Wang MB, Kim W, Yong WH, Heaney AP: **A human ACTH-secreting corticotroph tumoroid model: Novel Human ACTH-Secreting Tumor Cell in vitro Model**. *EBioMedicine* 2021, **66**:103294.

29. Feng Y, Mao ZG, Wang X, Du Q, Jian M, Zhu D, Xiao Z, Wang HJ, Zhu YH: **MicroRNAs and Target Genes in Pituitary Adenomas**. *Horm Metab Res* 2018, **50**(3):179-192.

30. Cheunsuchon P, Zhou Y, Zhang X, Lee H, Chen W, Nakayama Y, Rice KA, Tessa Hedley-Whyte E, Swearingen B, Klibanski A: **Silencing of the imprinted DLK1-MEG3 locus in human clinically nonfunctioning pituitary adenomas**. *Am J Pathol* 2011, **179**(4):2120-2130.

31. Vicchio TM, Aliquò F, Ruggeri RM, Ragonese M, Giuffrida G, Cotta OR, Spagnolo F, Torre ML, Alibrandi A, Asmundo A *et al*: **MicroRNAs expression in pituitary tumors: differences related to functional status, pathological features, and clinical behavior**. *J Endocrinol Invest* 2020, **43**(7):947-958.

32. Amaral FC, Torres N, Saggioro F, Neder L, Machado HR, Silva WA, Jr., Moreira AC, Castro M: **MicroRNAs differentially expressed in ACTH-secreting pituitary tumors**. *J Clin Endocrinol Metab* 2009, **94**(1):320-323.

33. Gentilin E, Tagliati F, Filieri C, Molè D, Minoia M, Rosaria Ambrosio M, Degli Uberti EC, Zatelli MC: **miR-26a plays an important role in cell cycle regulation in ACTH-secreting pituitary adenomas by modulating protein kinase Cδ**. *Endocrinology* 2013, **154**(5):1690-1700.

34. Varghese RS, Barefoot ME, Jain S, Chen Y, Zhang Y, Alley A, Kroemer AH, Tadesse MG, Kumar D, Sherif ZA *et al*: **Integrative Analysis of DNA Methylation and microRNA Expression Reveals Mechanisms of Racial Heterogeneity in Hepatocellular Carcinoma**. *Front Genet* 2021, **12**:708326.

35. Li J, Huang HY, Lin YC, Zuo H, Tang Y, Huang HD: **Cinnamomi ramulus inhibits cancer cells growth by inducing G2/M arrest**. *Front Pharmacol* 2023, **14**:1121799.

36. Lundberg SM, Lee S-I: **A Unified Approach to Interpreting Model Predictions**. 2017:4765--4774.

37. Shao X, Yu L, Li C, Qian J, Yang X, Yang H, Liao J, Fan X, Xu X, Fan X: **Extracellular vesicle-derived miRNA-mediated cell–cell communication inference for single-cell transcriptomic data with miRTalk**. *Genome Biology* 2025, **26**(1):95.

38. Guessous F, Alvarado-Velez M, Marcinkiewicz L, Zhang Y, Kim J, Heister S, Kefas B, Godlewski J, Schiff D, Purow B *et al*: **Oncogenic effects of miR-10b in glioblastoma stem cells**. *J Neurooncol* 2013, **112**(2):153-163.

39. Tesfaye AA, Azmi AS, Philip PA: **miRNA and Gene Expression in Pancreatic Ductal Adenocarcinoma**. *Am J Pathol* 2019, **189**(1):58-70.

40. Huang H-Y, Lin Y-C-D, Cui S, Huang Y, Tang Y, Xu J, Bao J, Li Y, Wen J, Zuo H *et al*: **miRTarBase update 2022: an informative resource for experimentally validated miRNA–target interactions**. *Nucleic Acids Research* 2022, **50**(D1):D222-D230.

41. Ragab EM, El Gamal DM, Mohamed TM, Khamis AA: **Therapeutic potential of chrysin nanoparticle-mediation inhibition of succinate dehydrogenase and ubiquinone oxidoreductase in pancreatic and lung adenocarcinoma**. *European Journal of Medical Research*

2022, **27**(1):172.
42. Krol J, Loedige I, Filipowicz W: **The widespread regulation of microRNA biogenesis, function and decay**. *Nat Rev Genetics* 2010, **11**(9):597-610.
43. Vilimova M, Pfeffer S: **Post‐transcriptional regulation of polycistronic microRNAs**. *Wiley Interdiscip Rev Rna* 2022:e1749.
44. Goldman MJ, Craft B, Hastie M, Repečka K, McDade F, Kamath A, Banerjee A, Luo Y, Rogers D, Brooks AN *et al*: **Visualizing and interpreting cancer genomics data via the Xena platform**. *Nature Biotechnology* 2020, **38**(6):675-678.
45. Han X, Zhou Z, Fei L, Sun H, Wang R, Chen Y, Chen H, Wang J, Tang H, Ge W *et al*: **Construction of a human cell landscape at single-cell level**. *Nature* 2020, **581**(7808):303-309.
46. Ioffe S, Szegedy C: **Batch normalization: Accelerating deep network training by reducing internal covariate shift**. In: *International conference on machine learning: 2015*. PMLR: 448-456.
47. Srivastava N, Hinton G, Krizhevsky A, Sutskever I, Salakhutdinov R: **Dropout: A Simple Way to Prevent Neural Networks from Overfitting**. *J Mach Learn Res* 2014, **15**:1929-1958.
48. Agarap AF: **Deep learning using rectified linear units (relu)**. *arXiv preprint arXiv:180308375* 2018.
49. Kern F, Fehlmann T, Solomon J, Schwed L, Grammes N, Backes C, Van Keuren-Jensen K, Craig DW, Meese E, Keller A: **miEAA 2.0: integrating multi-species microRNA enrichment analysis and workflow management systems**. *Nucleic Acids Research* 2020, **48**(W1):W521-W528.
50. Bastian M, Heymann S, Jacomy M: **Gephi: An open source software for exploring and manipulating networks**. *Proceedings of the International AAAI Conference on Web and Social Media* 2009, **3**((1)):361-362.
51. Ashburner M, Ball CA, Blake JA, Botstein D, Butler H, Cherry JM, Davis AP, Dolinski K, Dwight SS, Eppig JT *et al*: **Gene ontology: tool for the unification of biology. The Gene Ontology Consortium**. *Nat Genet* 2000, **25**(1):25-29.
52. Wu T, Hu E, Xu S, Chen M, Guo P, Dai Z, Feng T, Zhou L, Tang W, Zhan L *et al*: **clusterProfiler 4.0: A universal enrichment tool for interpreting omics data**. *The Innovation* 2021, **2**(3).
53. Wong NW, Chen Y, Chen S, Wang X: **OncomiR: an online resource for exploring pan-cancer microRNA dysregulation**. *Bioinformatics* 2018, **34**(4):713-715.
54. Tang Z, Kang B, Li C, Chen T, Zhang Z: **GEPIA2: an enhanced web server for large-scale expression profiling and interactive analysis**. *Nucleic Acids Res* 2019, **47**(W1):W556-W560.
55. Love MI, Huber W, Anders S: **Moderated estimation of fold change and dispersion for RNA-seq data with DESeq2**. *Genome Biology* 2014, **15**(12):550.
56. Hao Y, Hao S, Andersen-Nissen E, Mauck WM, 3rd, Zheng S, Butler A, Lee MJ, Wilk AJ, Darby C, Zager M *et al*: **Integrated analysis of multimodal single-cell data**. *Cell* 2021, **184**(13):3573-3587 e3529.

# Supplementary material

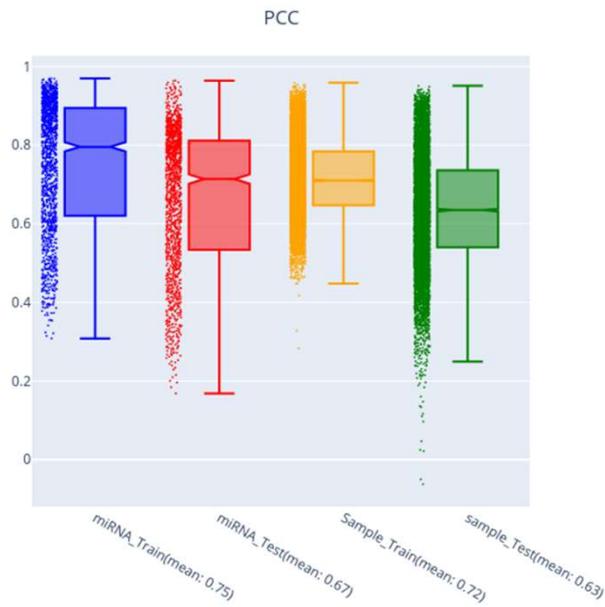

**Figure S1: Box plot for model performance by 3-fold cross-validation.** (Blue and red) Training and test dataset average PCC of miRNA among all samples. (Yellow and Green) Boxplot of average PCC of samples among all miRNAs in the training and test dataset, respectively.

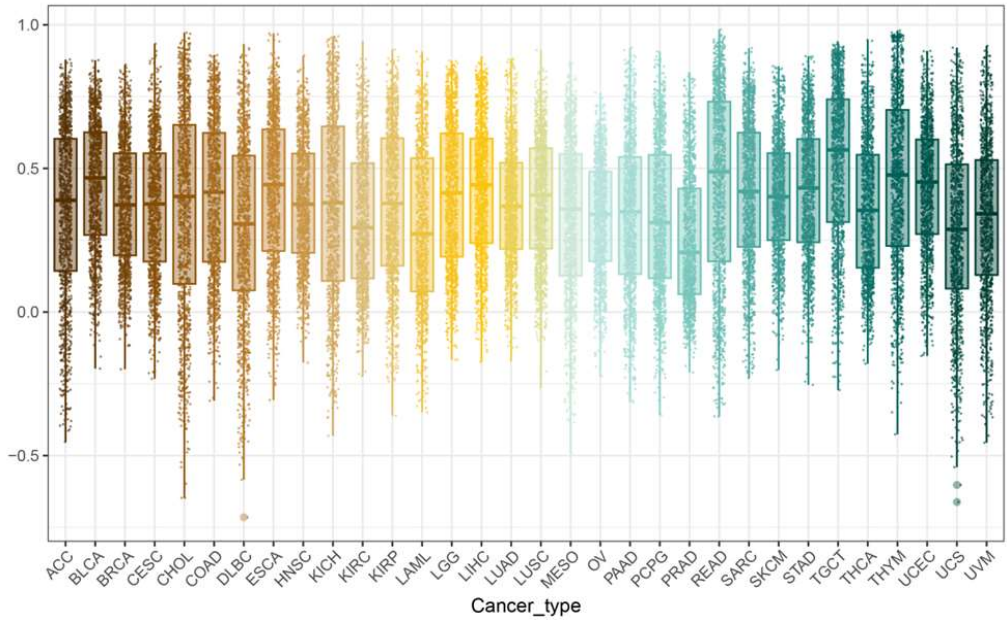

**Figure S2:** Average model performance (PCC) of each miRNA in different cancer types predicted by model trained with all types of cancers.

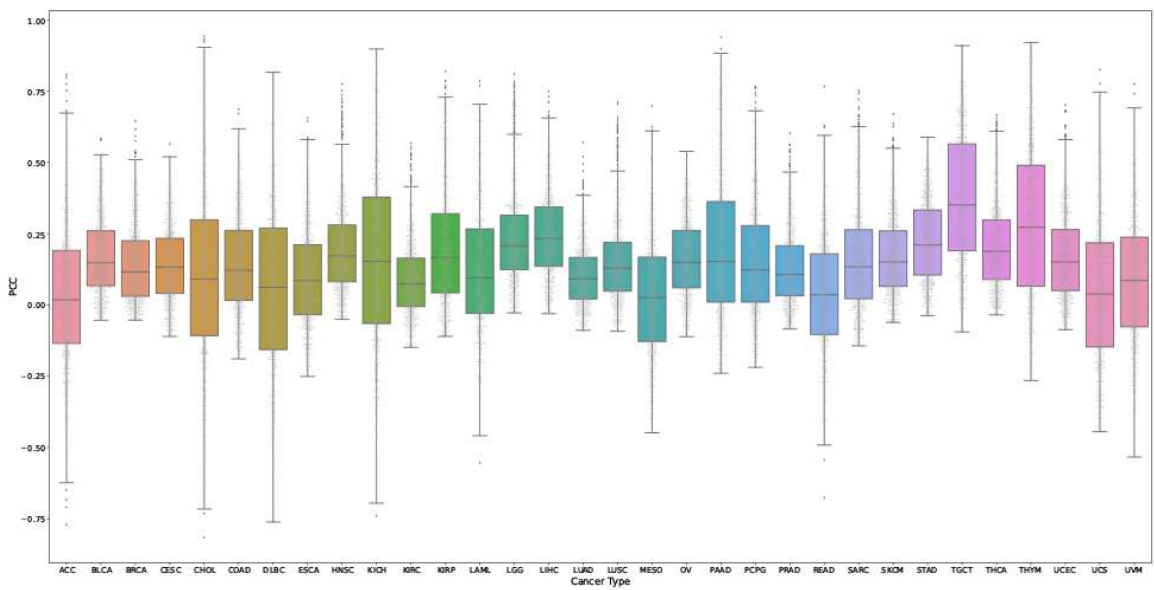

**Figure S3:** Box plot for performance of cancer-type specific models.

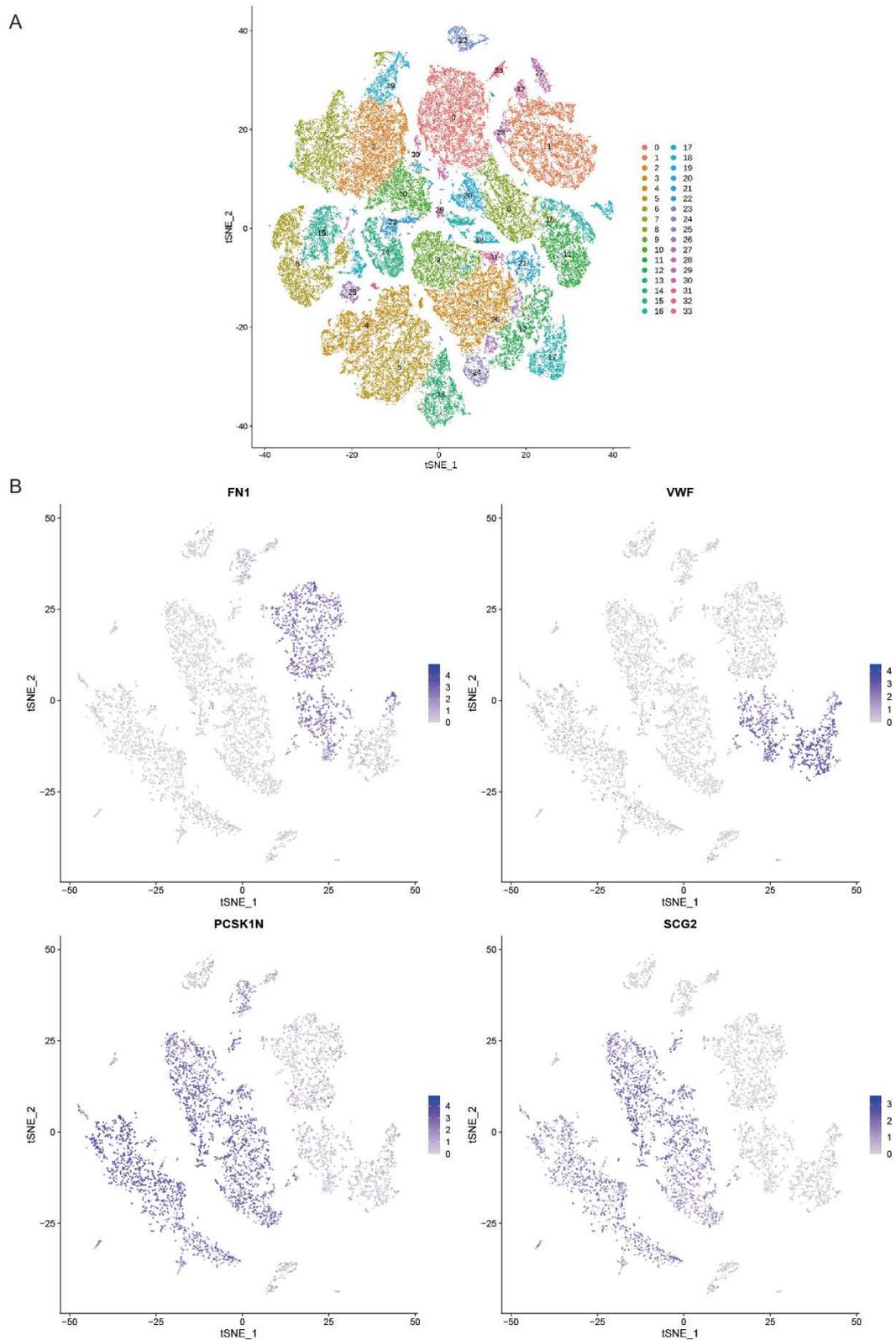

**Figure S4: T-SNE for scRNA-seq analysis. A** T-SNE distribution of different clusters of PDAC. **B** T-SNE distribution and cell-type marker expression among PA.

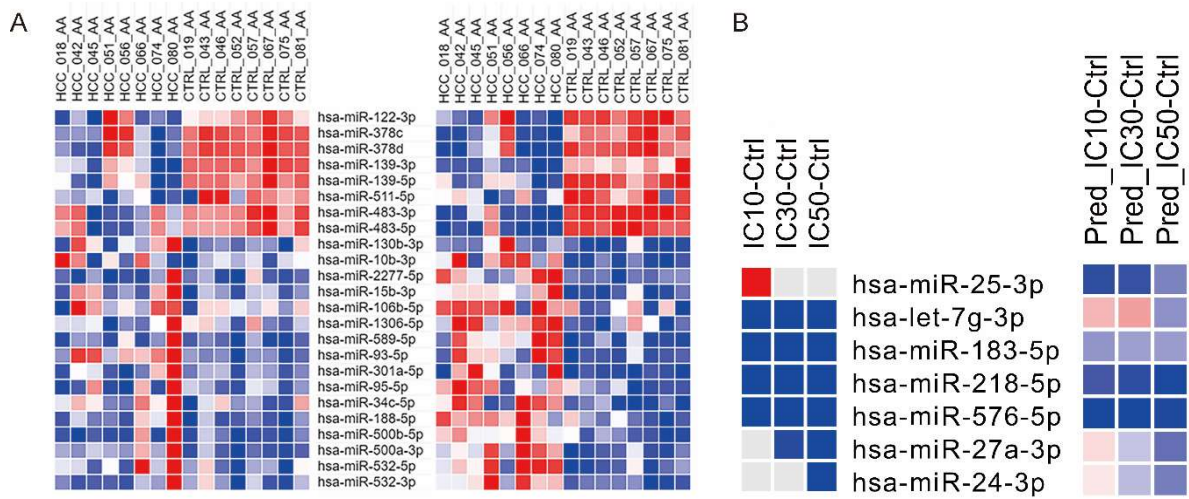

**Figure S5:** **Heatmap comparing predicted and real data of bulk RNA-seq for DEmiRs in A** cancerous and normal tissue for liver cancer. **B** TCM-perturbated A549 cell line.

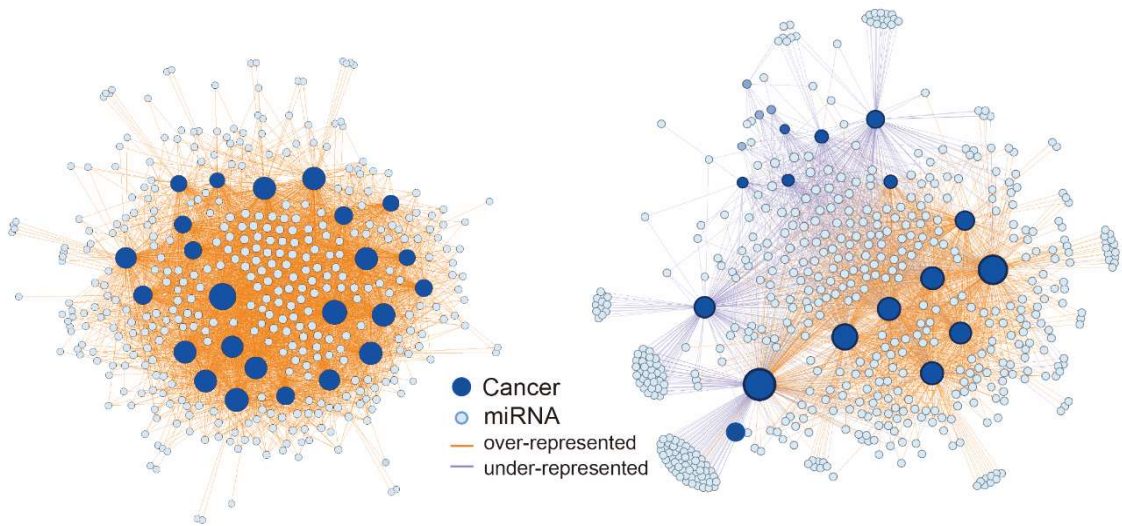
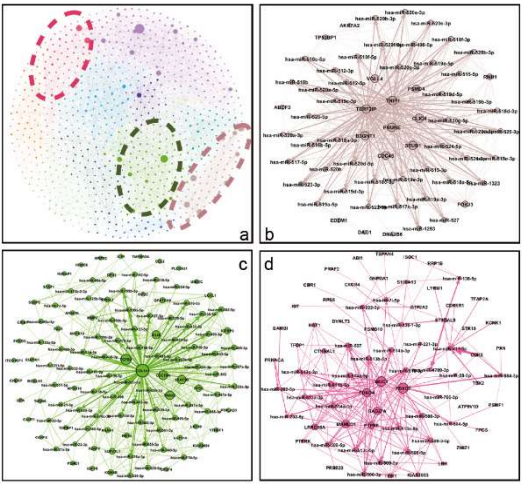
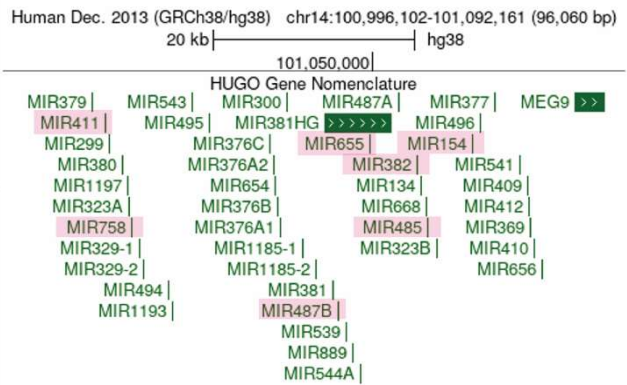
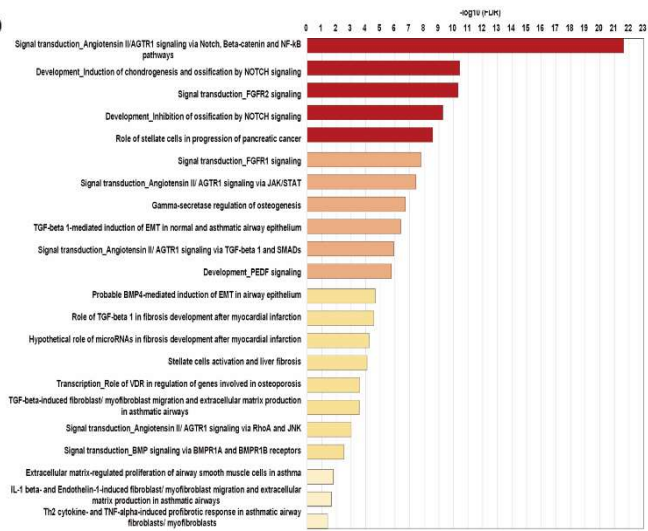

**Figure S6:** miRNAs with PCC >=0.8 have tighter association with cancer development,

**metastasis and prognosis. A** Reported associations between miRNAs and cancers annotated by miEAA 2.0 retrieving from HMDD database are demonstrated in networks. Network (left) shows the associations between cancers and miRNAs with PCC >=0.8. Network (right) shows the association between cancers and miRNAs with PCC <0.8. Over-represented of miRNAs in cancers means the miRNAs over-expressed in cancer tissues, vise-versus. **B** The contribution of landmark genes as features to expression of miRNAs are clustered into (a) 12 modules. (b-d)3 modules are visualized individually. **C** Host genes of miRNAs with whose expression positively contributed by COL1A1 locate densely at chromosome 14. **D** Enrichment analysis for Gene Ontology on target genes of miRNAs and contributing features.

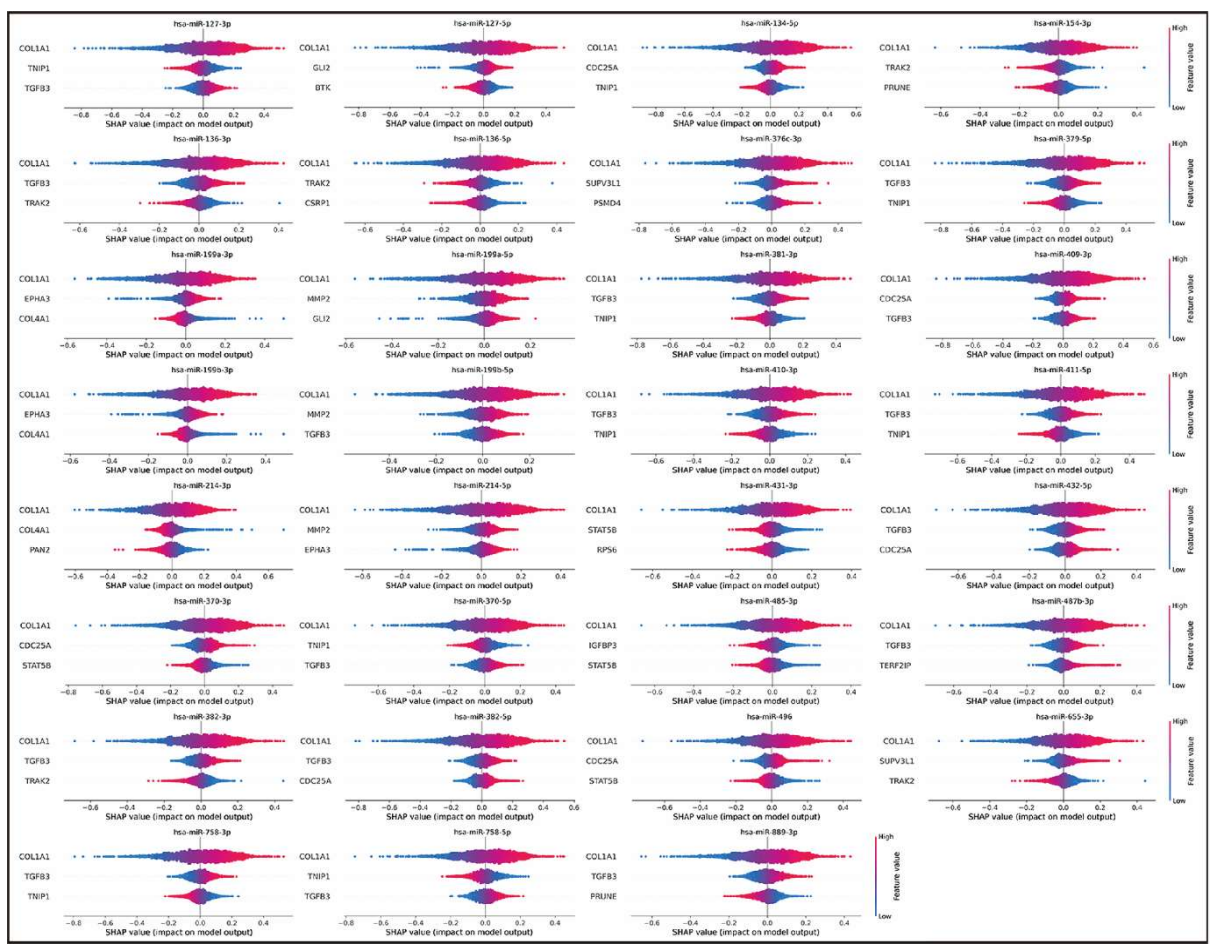

**Figure S7: SHAP analysis facilitated model interpretation and hub-miRNA discovery.**

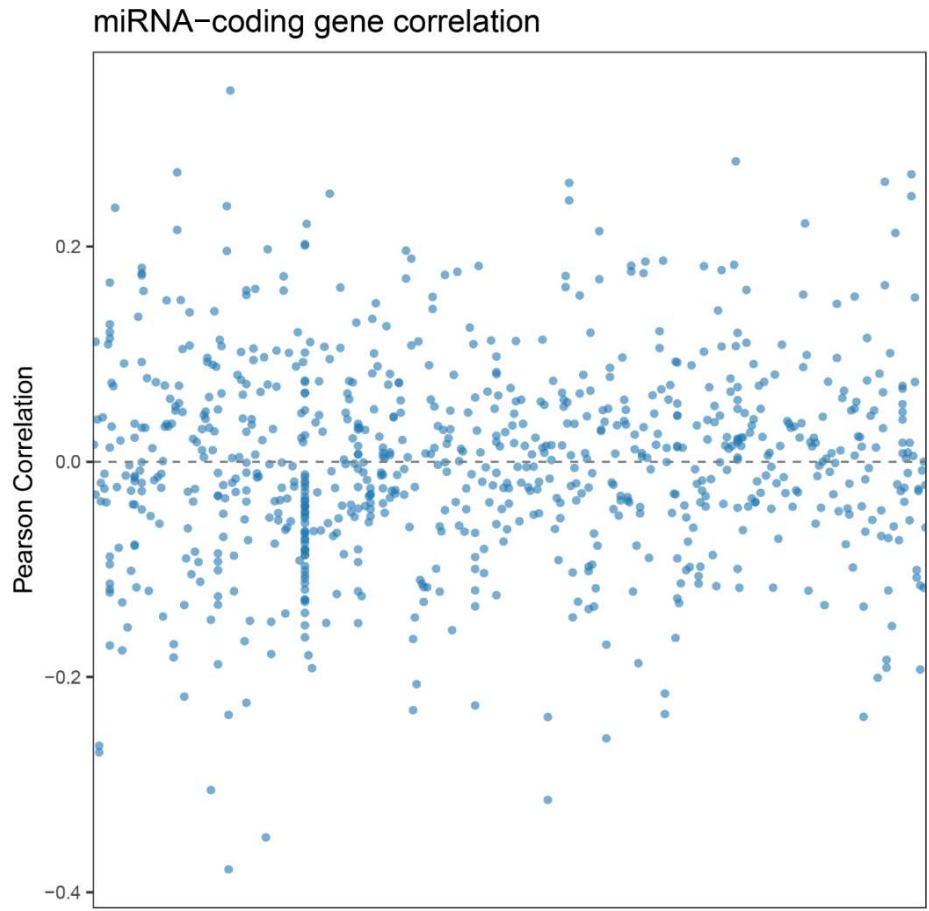

**Figure S8: Pearson Correlation Coefficient between mature miRNA expression and mRNA expression of miRNA coding genes in paired profiles from TCGA.** X-axis denotes miRNAs. Information of miRNA coding genes are retrieved from miRStart2.

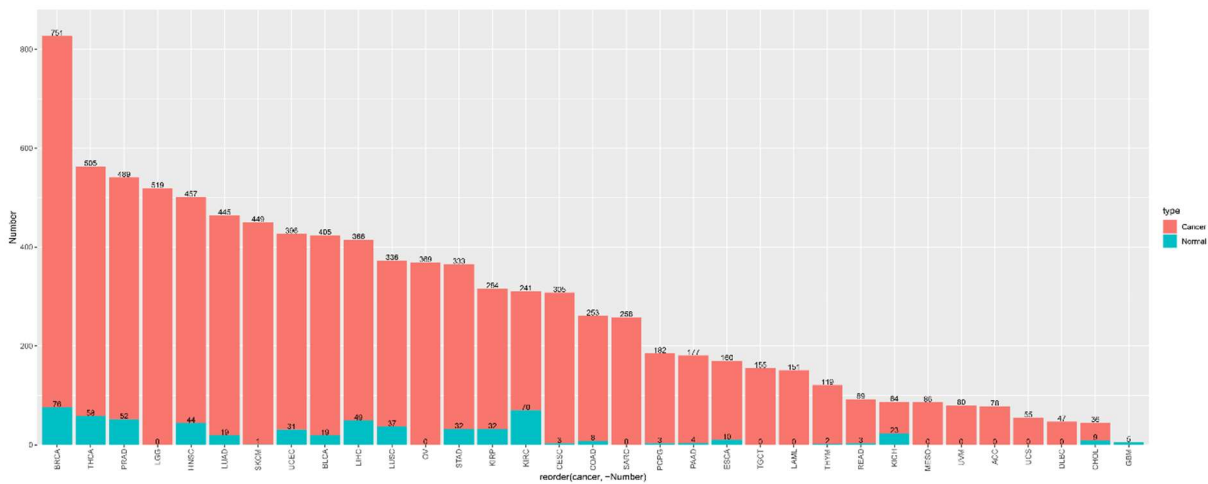

**Figure S9: Statistics of samples of cancers and corresponding normal tissues.**

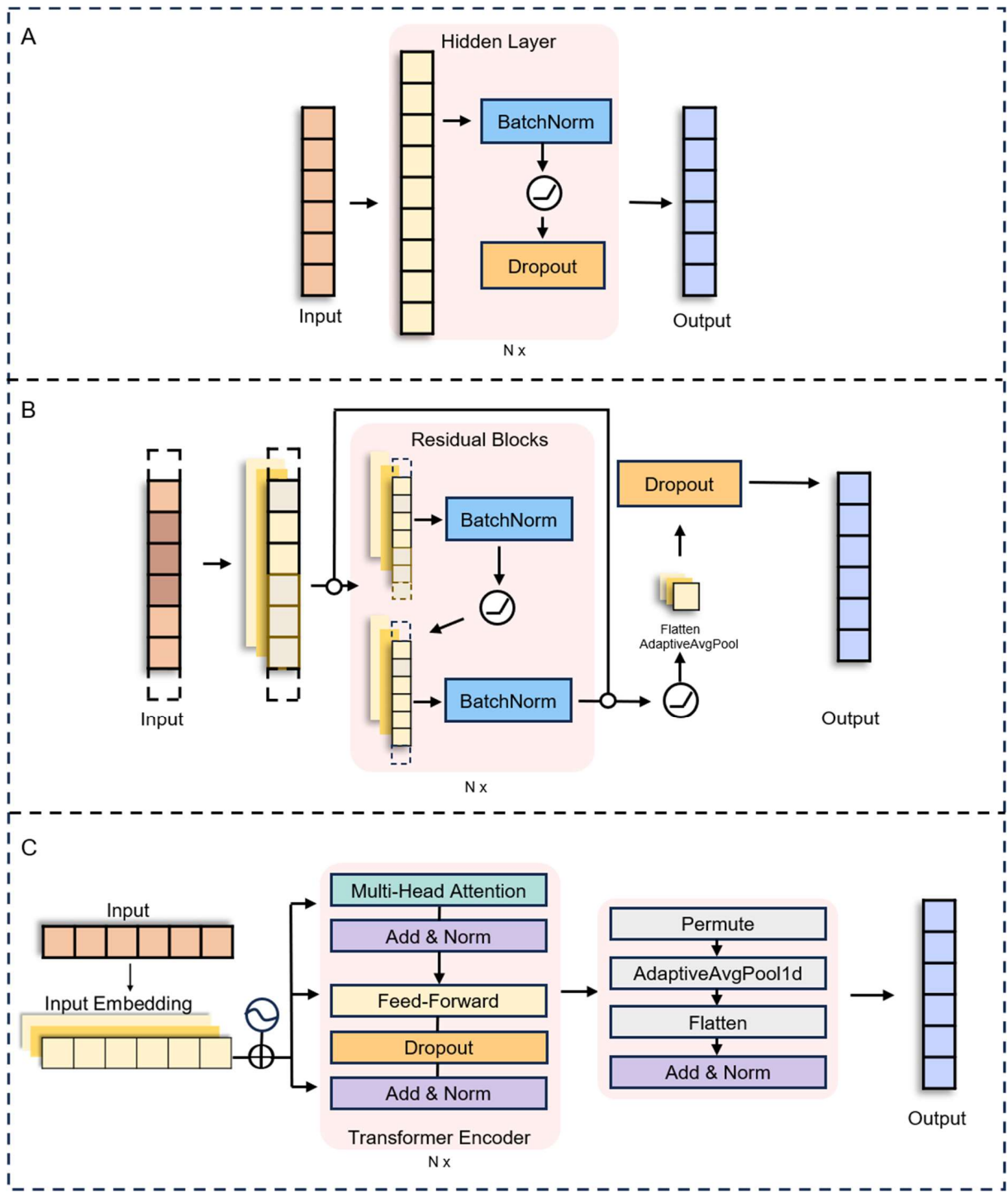

**Figure S10: Model architecture of models used for model constructure and comparison.**
**A** Model architecture of (deep) neural network. **B** Model architecture of ResNet. **C** Model architecture of Transformer.

**Supplementary File 1.xlsx:** Data for application of SiCmiR in miRNA expression prediction and potential hub-miRNA discovery in PDAC.

**Supplementary File 2.xlsx:** Data for application of SiCmiR in miRNA expression prediction and potential hub-miRNA discovery in ACTH-secreting tumor data.

**Supplementary File 3.xlsx:** Fold change of bulk sequenced and predicted miRNAs in liver cancers and TCM treated A549 cell line.

**Supplementary File 4.xlsx:** SHAP analysis, network analysis, and enrichment analysis data facilitate hub-miRNA discovery from the result of SiCmiR model.

**Supplementary File 5.xlsx:** EVmiR score, MTI score and specificity of significant miRNA-target pairs from sender and receiver cells.

**Supplementary File 6.xlsx:** Search space of hyperparameters for model training.